\documentstyle[12pt,emulateapj] {article}

\lefthead{EVANS ET AL.} 
\righthead{CO IN RADIO GALAXIES}

\begin{document}

\title{Molecular Gas and Nuclear Activity in Radio Galaxies Detected by {\it IRAS}}

\author{A. S. Evans\altaffilmark{1,2},
J. M.  Mazzarella\altaffilmark{3},
J. A. Surace\altaffilmark{4},
D. T. Frayer\altaffilmark{4},
K. Iwasawa\altaffilmark{5},
\& D. B. Sanders\altaffilmark{6,7}}

\altaffiltext{1}{Department of Physics \& Astronomy, Stony Brook
University, Stony Brook, NY, 11794-3800: aevans@mail.astro.sunysb.edu}

\altaffiltext{2}{Visiting Astronomer at the Infrared
Processing \& Analysis Center, California Institute of Technology, MS
100-22, Pasadena, CA 91125}

\altaffiltext{3}{Infrared
Processing \& Analysis Center, California Institute of Technology, MS
100-22, Pasadena, CA 91125: mazz@ipac.caltech.edu}

\altaffiltext{4}{SIRTF Science Center, California Institute of Technology,
Pasadena, CA 91125: jason@ipac.caltech.edu}

\altaffiltext{5}{Institute of Astronomy, Madingly Road, Cambridge CB3 0HA}

\altaffiltext{6}{Institute for Astronomy, 2680
Woodlawn Dr., Honolulu, HI 96822: sanders.ifa.hawaii.edu}

\altaffiltext{7}{Max-Plank Institut fur Extraterrestrische Physik,
D-85740, Garching, Germany}

\begin{abstract}

This paper reports the latest results from a millimeter-wave (CO)
spectroscopic survey of IRAS-detected radio galaxies with $L_{\rm 1.4GHz}
\sim 10^{23-28}$ W Hz$^{-1}$ in the redshift range $z \sim 0.02-0.15$.
The IRAS flux-limited sample contains 33 radio galaxies with different
radio morphologies and a broad range of infrared luminosities ($L_{\rm
IR} = 10^{9-12}$ L$_\odot$), allowing for an investigation of {\it (a)}
whether low-$z$ radio-selected AGN reside in molecular gas-rich host
galaxes, and {\it (b)} whether the CO properties are correlated with the
properties of the host galaxy or the AGN. All of the radio galaxies in
Mazzarella et al. (1993) and Mirabel et al. (1989) have been reobserved.
Three new CO detections have been made, raising the total number of CO
detections to nine and setting the survey detection rate at $\sim$ 25\%.
Many of the CO lines have double-peaked profiles, and the CO line widths
are broad (average $\Delta v_{\rm FWHM} \sim 500\pm130$ km s$^{-1}$),
exceeding the average CO widths of both ultraluminous infrared galaxies
($300\pm90$ km s$^{-1}$) and Palomar-Green QSOs ($260\pm160$ km s$^{-1}$),
and thus being indicative of massive host galaxies. The CO luminosities
translate into molecular gas masses of $\sim 0.4-7\times10^9$ M$_\odot$,
however, the 3$\sigma$ CO upper limits for nondetections do not rule out a
molecular gas mass as high as that of the Milky Way ($\sim 3\times10^9$
M$_\odot$).  Optical images of eight out of nine molecular gas-rich
radio galaxies show evidence of close companions and/or tidal features.
Finally, there is no obvious correlation between radio power and
molecular gas mass. However, it is notable that only one F-R II galaxy
out of 12 is detected in this CO survey; the remaining detections are
of galaxies hosting F-R I and compact radio jets.

\end{abstract}

\keywords{
galaxies: active ---
galaxies: interacting ---
galaxies: ISM ---
ISM: molecules ---
infrared: galaxies --
}

\section{Introduction}

A connection between galaxy-galaxy mergers and the onset of nuclear
activity in classical optical and radio-selected Active Galactic Nuclei
(AGN) has been inferred over the last 20 years.  Many classical radio
galaxies and QSOs have been observed at optical and near-infrared
wavelengths to have double-nuclei, tidal bridges/tails, and close
companions consistent with them being the mergers of disk galaxies (e.g.
Stockton \& MacKenty 1983; Heckman et al.  1986; Smith \& Heckman 1990a,b;
McLeod \& Rieke 1994; Surace, Sanders \& Evans 2001). Further evidence of
the AGN-merger connection has resulted from multiwavelength surveys of
luminous infrared-selected galaxies, which have shown these galaxies
to be merging/interacting disk galaxies and have shown an increasing
fraction of infrared galaxies with optical, near-infrared, and X-ray
signatures of imbedded AGN as a function of increasing luminosity (e.g.,
Sanders et al. 1988a,b; Veilleux, Sanders, \& Kim 1997, 1999; Veilleux et
al 1999; Vignati et al. 1999).  

Star formation appears to be a component of the activity occuring in
the host galaxies of AGN -- a growing number of radio galaxies and QSOs
have detectable amounts of molecular gas and young UV-luminous knots
indicative of recent star formation (Philips et al. 1987; Sanders,
Scoville, \& Soifer 1988c; Lazareff et al. 1989;  Mirabel, Sanders,
\& Kaz\`{e}s 1989; Mazzarella et al. 1993 -- hereafter, Paper 1;
Reuter et al.  1993; Evans et al.  2001; Allen et al.  2002; Scoville
et al.  2003).  In addition, the fraction of classical AGN with dust
in their host galaxies is substantial (e.g., Golombek et al. 1988; Sanders
et al 1989), with $\sim$ 50\% of 3C radio galaxies and $\sim$ 70\% of
Palomar-Green (PG: Schmidt \& Green 1983) QSOs at $z \lesssim 0.12$ having
IRAS\footnote{I.e., the Infrared Astronomical Satellite.} detections
at 60$\mu$m.  The Infrared Space Observatory (ISO) detected thermal dust
emission in a larger fraction of radio galaxies and QSOs (Meisenheimer et
al. 2001; Haas et al. 2003, 2004), providing further proof of rich ISMs in
the host galaxies of radio galaxies and QSOs.  This dust is most likely
a combination of dust intrinsic to the progenitor galaxies, as well as
dust produced through stellar mass loss from a recent/ongoing starburst.

The importance of star formation in AGN host galaxies is also inferred via
recent black hole surveys in nearby galaxies.  These observations
have shown that the massive spheroidal component of many nearby normal
galaxies contain quiescent supermassive nuclear black holes ($M_\bullet
= 10^{6-9}$ M$_\odot$: e.g. Magorrian et al. 1998).  A key result that
has come from these surveys is the scaling of black hole mass with
stellar bulge mass (e.g., Magorrian et al. 1998). Such a correlation may
indicate that the formation of stars in the bulge component of galaxies
is intimately associated with the building of supermassive black holes.

While the molecular gas properties of infrared-selected galaxies have been
studied in detail, CO observations of large samples of radio galaxies
and QSOs have lagged behind. Thus, conclusions about the CO properties
of classical AGN and their host galaxies is often based on small number
statistics or incomplete data sets.  The present study focusses on
the global molecular gas properties of a sample of low-redshift radio
galaxies, and it serves as an extension of the survey presented in Paper
1. The primary issues to be addressed with this survey are {\it (a)}
whether low-$z$ radio-selected AGN reside in molecular gas-rich host
galaxies and {\it (b)} whether the CO properties are correlated with the
properties of the host galaxy or the AGN.  To address the latter issue,
the millimeter data are complemented by infrared, radio, and X-ray data to
allow a comparison between the available amount of fuel for star formation
(and perhaps AGN activity) and the energy emitted from the AGN and/or
host galaxy at these wavelengths. The analysis presented here builds
primarily upon similar analysis done of low-redshift, infrared-selected
galaxies over the last twenty years (e.g., Sanders \& Mirabel 1996 and
references therein; Franceschini et al. 2003; Trentham et al. 2005)
and recent analysis of a sample of CO-detected PG QSO host galaxies
(Evans et al. 2001). These galaxies, which may emanate from the same
parent population as the radio galaxies or be linked via evolution, will
serve as the primary comparison samples.  In addition, optical imaging
of a radio galaxy subsample will be briefly discussed. Parts of this
extended millimeter survey have been presented elsewhere (Evans 1996;
Sanders \& Mirabel 1996; Evans 1998; Evans et al. 1999a,b).

This paper is divided into six sections. The selection criteria of the
radio galaxy sample is discussed in \S 2. Section 3 is a summary of the
observations and data reduction. In \S 4, the CO($1\to0$) emission-line
properties are presented, along with the method of calculating the
molecular gas mass. Section 5 contains a brief summary of the optical
imaging data and an extended discussion of the infrared,
radio, CO, and X-ray properties of the radio galaxies relative to those of
infrared-selected galaxies, PG QSOs, and elliptical galaxies. Section 6
is a summary. Throughout the paper, $H_0 = 75$ km s$^{-1}$ Mpc$^{-1}$,
$q_0 = 0.5$, and $\Lambda = 0.0$ are assumed.

\section{Sample}

The radio galaxy sample was selected primarily from a list of radio
galaxies correlated with the IRAS archive by Golombek, Miley \& Neugebauer
(1988).  The radio galaxies were selected for CO observations based on
the following criteria: {\it (i)} a declination, $\delta > -20^{\rm
o}$, to allow observations from both the Kitt Peak 12m Telescope
and the James Clerk Maxwell Telescope (JCMT), {\it (ii)} {\it IRAS}
flux densities at 60$\micron$, $f_{\rm 60\mu m}$, or 100$\micron$,
$f_{\rm 100\mu m}$, greater than $0.3$ Jy, with the assumption that
an {\it IRAS} detection at this level is a good indicator of a dusty,
gas-rich interstellar medium, and {\it (iii)} a redshift in the range
$0.02 \lesssim z \lesssim 0.15$, to ensure detectability of the molecular
gas mass of the Milky Way ($\sim 3\times10^9$ M$_\odot$), if present,
with single-dish millimeter telescopes used for the survey (for the
sensitivity calculation, the assumption is made that the CO lines have a
full width at half the maximum intensity width of $\Delta v_{\rm FWHM}
\sim 250$ km s$^{-1}$, which is the average value for a sample of IRAS
galaxies observed by Sanders, Scoville, \& Soifer 1991). Such criteria
yield 31 radio galaxies. Two additional radio galaxies, PKS 0624-20 and
3C 285, were identified via a search of the NASA Extragalactic Database
(NED) to have properties consistent with the selection criteria; their
addition to the sample raises the total number of galaxies to 33.
Note that two of the 33 galaxies, the relatively nearby galaxies NGC
6251 and 3C 272.1, have CO non-detections in published surveys by Elfhag
et al. (1996) and Knapp \& Rupen (1996), and thus were not reobserved
as part of this survey.  General properties of the sample of 33 radio
galaxies are provided in Table 1.

\section{Observations and Data Reduction}

\subsection{CO Spectroscopy}

Millimeter observations of the redshifted CO($1\to0$) emission in
31 of the 33 radio galaxies were made with the NRAO\footnote{The
NRAO is a facility of the National Science Foundation operated under
cooperative agreement by Associated Universities, Inc.} 12m Telescope
during six observing periods between 1992 December and 2000 June.
The telescope was configured with two $256\times2$ MHz channel filterbanks
and dual polarization SIS spectral-line receivers tuned to the frequency
corresponding to the redshifted CO($1\to0$) emission line. (Optical
emission line redshifts were adopted.) Observations were obtained using
a nutating subreflector with a chop rate of $\sim 1.25$ Hz. Six minute
scans were taken, and a calibration was done every other scan. During
the course of each observation, data were also taken with the bandpass
centered at velocity offsets of $\pm$ 20 or 50 km s$^{-1}$ of the velocity
corresponding to the redshift of the optical emission lines. Shifting
the velocities in such a manner minimizes any ripples inherent in the
baseline. Pointing was checked every few hours by observing standard
continuum sources and was generally determined to be accurate to within
a few arcseconds.

Observations of the redshifted CO($1\to0$) emission in Cygnus A
and PKS 1345+12 were done at the IRAM 30m telescope during 1995
December and 2003 December observing periods, respectively. The 3 and
1 millimeter receivers were used simultaneously in combination with the
filterbanks and an autocorrelator to provide bandwidths of 512 and 600 MHz,
respectively. During observations, the pointing was monitored by observing
the planets and standard continuum sources.

Observations of the redshifted HCN($1\to0$) emission in PKS 0502-10 and
B2 0722+30 were done at the IRAM 30m telescope during the 2003 December
observing period. The setup was identical to the CO($1\to0$) observation
of PKS 1345+12 obtained during the same period. HCN observations of
the strong 3mm continuum galaxies 3C 84 and 3C 120 were also attempted,
but have been left out of the present discussion due to the presence of
large baseline ripples in the data.

Observations of the redshifted CO($2\to1$) emission in PKS 0502-10
and B2 0722+30 were done with the James Clerk Maxwell Telescope (JCMT)
in 1995 September.  The spectral line receiver (A2) were used together
with the Digital Autocorrelation Spectrometer in wide-band mode (750 MHz
bandwidth). Because of excess noise near the edge of the passband, the
usable bandwidth was only $\sim$ 700 MHz, which corresponds to a total
velocity coverage of $\sim$ 900 km s$^{-1}$ at 230 GHz.  All observations
were obtained using a nutating subreflector with a chop rate of $\sim$
1.25 Hz. Data were stored as six-minute scans, and a chopped wheel
calibration was performed after every other scan. Pointing was monitored
every few hours, and velocity shifts of $\pm$ 20 km s$^{-1}$ were made
to minimize baseline ripples. A journal of observations for all of the
single-dish observations are provided in Table 2.

The data reduction for all of the single-dish millimeter and submillimeter
observations were reduced using the IRAM data reduction package
CLASS.  The individual scans were checked for baseline instabilities, then
averaged together.  A linear baseline was subtracted from each spectrum;
the cases where emission lines were detected, the baseline subtraction
was done at velocities outside of the range of the emission line.
Each spectrum was smoothed to $\sim 20-50$ km s$^{-1}$. Finally, line
fluxes were measured by numerically integrating over the channels
in the line profile, and the line widths were measureed as full width
at 50\% of the peak flux. The resultant
spectra are plotted in Figures 1--3.

\subsection{Optical Imaging}

Ground-based imaging observations of 7 of the radio galaxies in the sample
were made at the UH 2.2m Telescope.  The U$^{\prime}$, B, and I-band
images were obtained over the period 1998 March 25 to 2000 December 31
using the Orbit Semiconductor 2048$\times$2048 CCD
camera. The original scale is $0.09\arcsec$/pixels (Wainscoat 1996),
but the CCD was read out with $2\times2$ pixel binning. Three to four
dithered exposures were taken, with integration times ranging from 360
to 720 seconds each for U$^{\prime}$, B, and I-band filters.

The U$^{\prime}$, B, and I-band data reduction was performed using IRAF.
The data reduction consisted of flatfielding individual images, scaling
each image to its median value to correct for offsets in individual
images, then shifting and median combining the images.  The final images,
boxcar smoothed by $4\times4$ pixels, are shown in Figure 4. 

\section{Results}


Table 3 is a summary of the CO emission-line properties of the 31 radio
galaxies observed as part of this survey. All of the radio galaxies
observed for Paper 1 were reobserved, as well as two additional radio
galaxies previously detected in CO (3C 84: Lazareff et al. 1989; Mirabel
et al. 1989; Reuter et al. 1993, PKS 1345+12: Mirabel et al. 1989). In
one case (TXS 1506+345), the pointing was shifted by $\sim 30\arcsec$
from that used in Paper 1 in order to center the galaxy in the beam. Nine
radio galaxies are detected in CO($1\to0$), including three new detections
(3C 31, PKS 0502-10, 3C 293)\footnote{Note that a detection of 3C 31
has also been reported by Lim et al. (2000).}.  However, the previously
reported detection of CO emission in B2 0648+27 was not confirmed. The
radio galaxy PKS 1345+12 was the weakest CO detection in the sample,
thus independent CO observations of the galaxy were also obtained with
the IRAM 30m Telescope to improve the determination of the CO flux and
line shape (Figure 1b).  The average full width at half the maximum
intensity velocity, $\Delta v_{\rm FWHM}$, of the detected galaxies is
500 km s$^{-1}$. Several of the galaxies have double-peaked profiles.

Table 3 lists the measured CO intensities and luminosities. The CO
luminosities of 3C 120, B2 0722+30, B2 1318+34, and TXS 1506+345 agree
with the values in Paper 1 to within $\sim$ 20\%, and the values of
$\Delta v_{\rm FWHM}$ for all except TXS 1506+345 are similar (TXS
1506+345 is reported as having a 2-component CO emission line profile in
Paper 1). There is a major discrepancy between the new CO measurements
of Perseus A and PKS 1345+12 and those in the literature. For Perseus,
the line width of CO luminosity are in agreement with the measurements
of Lazareff et al. 1989 and Reuter et al. (1993), but $L'_{\rm CO}$ is a
factor of two higher than the value determined by Mirabel et al. (1989).
The discrepancy with the Mirabel et al. (1989) measurements results from
what appears to be extended emission in the wings of the line in Figure
1 -- if this emission is ignored, then the CO luminosity is $6\times10^8$
K km s$^{-1}$ pc$^2$, within 22\% of the value reported by Mirabel et
al. (1989). The PKS 1345+12 CO luminosity comparison is more difficult
to rectify -- the CO luminosity is a factor of two lower than measured
by Mirabel et al. (1989). Note, however, that the errors associated with
the Mirabel et al. (1989) data are not known. The present PKS 1345+12 CO
measurements are correct to within the error of those reported in Evans
et al. (1999) if a 20\% calibrated error on the present observations
is assumed.

Many of the observations resulted in non-detections. For each of these
galaxies, upper limits to the CO intensity were calculated via,

$${T_{\rm mb} \Delta v} < {{3 T_{\rm rms} \Delta v_{\rm FWHM}} \over
{\sqrt {\Delta v_{\rm FWHM} / \Delta v_{\rm res}}}}\
~~~[{\rm K~km~s}^{-1}], \eqno(1)$$

\noindent
where $T_{\rm mb}$ is the main beam temperature, $\Delta v_{\rm FWHM}$ is
the average FWHM velocity for the
detected radio galaxies (= 500 km s$^{-1}$), and $T_{\rm rms}$ is the
root-mean-squared main beam temperature of the spectral data for a velocity
resolution of $\Delta v_{\rm res}$.

To calculate the CO luminosity, the luminosity distance for a source
at a given redshift and a $\Lambda = 0$ Universe was first calculated
using the equation

$$D_{\rm L} =
cH^{-1}_0q^{-2}_0 \left\{ z q_0 + (q_0 - 1) \left( \sqrt{2 q_0 z + 1} - 1
\right) \right\}$$
$$ ~[{\rm Mpc}].  \eqno(2)$$

\noindent
Given the measured CO flux, $S_{\rm CO} \Delta v$ [Jy km s$^{-1}$], the CO
luminosity of a source at redshift $z$ is

$$L'_{\rm CO} = \left( {c^2 \over {2 k \nu^2_{\rm obs}}} \right) S_{\rm
CO} \Delta v D^2_{\rm L} (1 + z)^{-3}$$
$$ ~[{\rm K~km~s}^{-1} {\rm~pc}^2],
\eqno(3)$$

\noindent
(Solomon, Downes, \& Radford 1992) where $c$ [km s$^{-1}$] is the speed of
light, $k$ [J K$^{-1}$] is the Boltzmann constant, and $\nu_{\rm obs}$
[Hz] is the observed frequency.  In terms of useful units, $L'_{\rm
CO(1\to0)}$ can be written as

$$L'_{\rm CO} = 2.4\times10^3 \left( S_{\rm CO} \Delta v
\over {\rm Jy~km~s}^{-1} \right) \left( D_{\rm L} \over {\rm Mpc}
\right) ^2 (1 + z)^{-1}$$
$$[{\rm K~km~s}^{-1} {\rm~pc}^2]. \eqno(4)$$

To estimate the mass of molecular gas in these radio galaxies, the
reasonable assumption that the CO emission is optically thick and
thermalized, and that it originates in gravitationally bound molecular
clouds, is made. Thus, the ratio of the H$_2$ mass and the CO
luminosity is given by

$$\alpha = {M({\rm H}_2) \over L^\prime_{\rm CO}} \propto {\sqrt {n({\rm
H}_2)} \over T_{\rm b}}$$
$$ ~~[{\rm M}_\odot ({\rm K~km~s}^{-1} {\rm
~pc}^2)^{-1}], \eqno(5)$$

\noindent
where $n($H$_2)$ and $T_{\rm b}$ are the density of H$_2$ and brightness
temperature for the CO(1$\to$0) transition (Scoville \& Sanders 1987;
Solomon, Downes, \& Radford 1992).  Multitransition CO surveys of
molecular clouds in the Milky Way (e.g. Sanders et al. 1993), and in
nearby starburst galaxies (e.g. G\"{u}sten et al. 1993) have shown that
hotter clouds tend to be denser such that the density and temperature
dependencies cancel each other. The variation in the value of $\alpha$
is approximately a factor of 2 for a wide range of kinetic temperatures,
gas densities, and CO abundance (e.g. $\alpha = 2-5 M_{\odot}$ [K km
s$^{-1}$ pc$^2]^{-1}$:  Radford, Solomon, \& Downes 1991).  Solomon et
al. (1997) and Downes \& Solomon (1998) have made use of dynamical mass
estimates of a low-redshift infrared galaxy sample observed in CO with
the Plateau de Bure Interferometer to argue that $\alpha$ may, in some
cases, be as low as 1 $M_{\odot}$ (K km s$^{-1}$ pc$^2)^{-1}$.

For the present sample of galaxies, an $\alpha = 1.5 M_{\odot}$ (K km
s$^{-1}$ pc$^2)^{-1}$ is adopted, which is the approximate value derived
for luminous and ultraluminous infrared galaxies (Solomon et al. 1997;
Downes \& Solomon 1998; Evans et al. 2002). However, note that
$\alpha$ for some, or all, of these galaxies may be a factor of
$2-3$ higher (i.e., the approximate value derived for bulk of the
molecular gas in the disk of the Milky Way: Scoville \& Sanders 1987;
Strong et al. 1988). The derived molecular gas masses are listed in
Table 3. Taking these masses and comparing them with the dust masses,
$M_{\rm dust}$, which are calculated via

$$M_{\rm dust} \sim 4.78 \left( {f_{\rm 100\mu m} \over {\rm Jy}} \right)
\left( { D_L \over {\rm Mpc} } \right)^2
\left( e^{143.88 / T_{\rm dust} } - 1 \right)$$
$$~ [{\rm M}_\odot] \eqno(6)$$

\noindent
(see Young et al. 1989) where the dust temperature, $T_{\rm dust}$
is calculated from the 60$\micron$ and 100$\micron$ flux densities via

$$T_{\rm dust} = -(1+z) \left[{82 \over \ln (0.3f_{60\mu{\rm m}}/f_{100\mu{\rm
m}})} - 0.5 \right]$$
$$ ~[{\rm K}], \eqno(7)$$

\noindent
yields an average molecular gas-to-dust mass ratio of $860\pm460$
(see Table 4).  By comparison, the gas-to-dust ratio is $\sim 600$
($\alpha = 4.8 M_{\odot}$ [K km s$^{-1}$ pc$^2]^{-1}$) for IRAS-detected
spiral and luminous infrared galaxies (Young \& Scoville 1991 and
references therein).  Note that the estimated dust masses in Table 4 are
likely lower limits; the bulk of the dust in these galaxies may be in
a colder ($< 30$K) dust component (Young \& Scoville 1991; Devereux \&
Young 1990).

Given the adopted value of $\alpha$, it is worthwhile to make a
conservative estimate of the dynamical mass in order to determine the
percentage of the total mass in the form of molecular gas.  The dynamical
mass within the region of the galaxies containing molecular gas and dust
can be calculated via

$$M_{\rm dyn} \sim {\Delta v^2_{\rm FWHM}
R_{\rm CO} \over G}
= 226 \left( \Delta v_{\rm FWHM} \over {\rm km~~s^{-1}} \right) ^2
\left( R \over {\rm pc} \right)$$
$$~~[{\rm M}_\odot], \eqno(8)$$

\noindent
where $\Delta v_{\rm FWHM}$ is the full CO velocity width at half
the maximum flux density (Table 3) and $R_{\rm CO}$ is the radius
of the CO distribution.  The quantity $R_{\rm CO}$ is estimated
by assuming optically thick, thermalized gas with a unity filling
factor and a blackbody temperature, $T_{\rm bb}$, equal to that of
the dust.  
Thus, the radius of the CO distribution is calculated via

$$R_{\rm CO} = \sqrt{L'_{\rm CO} \over \pi {T_{\rm bb} \Delta v_{\rm
FWHM}} }
 ~[{\rm pc}] \eqno(9)$$

\noindent (see Table 4).  The resultant size estimates yield $M_{\rm dyn}
\sim 0.22-2.5\times10^{10}$ M$_\odot$ and thus $M({\rm H}_2) / M_{\rm dyn}
= 0.10-0.94$ with a mean value
of $0.45\pm0.29$. 

Note that the values of $R_{\rm CO}$ are likely to be underestimated,
and thus the derived values of $M({\rm H}_2) / M_{\rm dyn}$ are lower
limits. For example, the column densities derived from $R_{\rm CO}$ and
$M$(H$_2$) are in the range $1-4\times10^{24}$ cm$^{-2}$. In contrast,
the X-ray column densities estimated from X-ray observations of 3C
120 and PKS 1345+12 are in the range of $0.2-4\times10^{22}$ cm$^{-2}$
(Sambruna et al. 1999; O'dea et al. 2000).  Further, the CO disks in
3C 293 (Evans et al. 1999a) and 3C 31 (Okuda et al. 2005) are extended
$\sim$ 6 kpc and 1 kpc, respectively, which are 6--12 times the sizes of
the respective CO distributions listed in Table 4. This issue will be
addressed further in a follow-up paper containing CO interferometric
maps of the sample (Evans et al. 2005a).

\section{Discussion}

The present CO survey has increased the number of radio galaxies
observed in Paper 1 by a factor of four. 
Three new CO detections have been made, yielding
a 25\% detection efficiency for the sample with the 12m telescope. The
CO detections translate into molecular gas masses of $0.4-7\times10^9$
M$_\odot$, with 3$\sigma$ upper limits for nondetections in the range of
$< 4\times10^8$ M$_\odot$ for a $z \sim 0.02$ galaxy to $<3.3\times10^9$
M$_\odot$ for a $z \sim 0.1$ galaxy. Thus, while only a quarter of the
sample was detected, the upper limits of the more distant members of
the sample do not rule out molecular gas mass as high as that of the Milky Way
($\sim 3\times10^9$ M$_\odot$).

There are two notable features in the CO spectra shown in Figure 1.  The
first is the broad linewidth -- the average $\Delta v_{\rm FWHM}$ of the present
sample of radio galaxies is $\sim 500\pm130$ km s$^{-1}$, which exceeds the
average value for both PG QSOs (260$\pm160$ km s$^{-1}$: Evans et al.
2005b) and ultraluminous infrared galaxies (300$\pm90$ km s$^{-1}$: Solomon
et al. 1997) detected in CO to date. This result likely indicates that the
host galaxies of molecular gas-rich radio galaxies are extremely massive.
The second notable feature is the double-peaked CO emission line profiles
in many of the spectra. Interferometric CO observations of two of these
galaxies are in the literature and show this feature to be present for (at
least) two different reasons. In the case of 3C 293, the ``double-peak''
appearance is caused by a CO absorption feature at the systemic velocity of
the galaxy (Evans et al.  1999a). In the case of 3C 31, the molecular gas
is distributed in a ring (Okuda et al. 2005).

Figure 4 shows optical images of 7 of the nine radio galaxies detected
in CO; two of the nine radio galaxies (B2 1318+34 and PKS 1345+12) are
well-studied systems that have been imaged by several groups (e.g. Sanders
et al. 1988b; Bushouse \& Stanford 1992; Surace et al. 1998; Evans et al.
1999b; Scoville et al. 2000; Kim, Veilleux, \& Sanders 2003),
and thus have not been reobserved by us. Most of these molecular
gas-rich galaxies have nearby companions (3C 31; PKS 1345+12) and/or
show clear evidence of being involved in a recent merger (3C 84; 3C 120;
PKS 0502-10; 3C 293; PKS 1345+12; B2 1318+34). The one exception is B2
0722+30, an edge-on spiral galaxy. While the images in Figure 4 support the
possibility that interaction and merger events may be a primary trigger
for radio-loud AGN activity, it is notable that many low-redshift radio galaxies
are elliptical galaxies (e.g., Sandage 1966; Martel et al. 1999).  Thus,
unless the elliptical galaxies are advanced mergers, other mechanisms
must play an important role in igniting the AGN activity observed in
these galaxies. 

Given the results of the present survey and recent CO surveys of
low-redshift PG QSOs (Evans et al. 2001; Scoville 2003), there is
sufficient data to begin a comparison of the molecular gas properties
and the global energy output (measured at infrared, radio, and X-ray
wavelengths) of classical UV and radio-selected AGN with infrared-selected
galaxies and elliptical galaxies. Before exploring such comparisons,
it is worth first considering two issues which have direct relevance to
the interpretation of the data. The first issue is the infrared bias
in the sample. While the sample was selected based on the 60$\mu$m
and 100$\mu$m flux densities, it is important to realize that {\it
(a)} the flux density limits are extremely low, and thus the 
infrared luminosity range of the sample spans several decades (i.e., $L_{\rm IR}
\sim 10^{9-12}$ L$_\odot$), and that {\it (b)} 55\%, 33\%, and 24\% of
$z \lesssim 0.1$ 3C, B2 and PKS galaxies, respectively, were detected
at 60$\mu$m by IRAS. The fraction for PG QSOs is even higher, with 69\%
of PG QSOs out to $z \sim 0.17$ (the redshift limit of the QSO CO survey)
having 60$\mu$m detections. Thus, infrared excesses in AGN host galaxies
from radio-selected (and UV-selected) surveys are quite common.

The second issue of relevance to the interpretation of the data is the
nature of the radio emission from the galaxies in the sample; i.e.,
whether the radio emission from all of the galaxies in the sample is due to
AGN jet emission, or whether some of the galaxies are actually nearby starburst
galaxies luminous enough in the radio to have been detected by the 3C,
B2 or PKS surveys. To address this issue, IRAS and 1.4 GHz data have been
compiled for the radio galaxies, along with complementary data sets for
infrared galaxies with both warm and cool 25-to-60$\mu$m colors (note that
classical AGN typically have warm colors, $f_{\rm 25\mu m} / f_{\rm 60\mu
m} > 0.2$), PG QSOs, and elliptical galaxies. Figure 5 shows the result
of the compilation; plotted is the 1.4 GHz-to-infrared luminosity ratio,
$L_{\rm 1.4GHz}$, versus the infrared luminosity, $L_{\rm IR}$. The
``radio-infrared'' correlation (see Condon 1992) is clearly visible in
this Figure; the radio-to-infrared ratio for infrared-selected galaxies
is relatively constant with increasing infrared luminosity. This has led
to speculation that the same population of massive stars responsible
for heating the dust is also responsible for the radio emission (through,
for example, the production of supernovae). Lonsdale, Smith, \& Lonsdale
(1995) showed that QSOs follow the same bolometric luminosity--radio
correlation as infrared galaxies.  Similarly, the QSOs follow the
infrared--radio correlation in Figure 5, indicating, perhaps, that a
significant fraction of their radio and infrared emission is associated
with massive stars, or that AGN-related processes in radio-quiet sources
yield the proper $L_{\rm 1.4GHz}/L_{\rm IR}$ ratio to place QSOs on
the relation.  In contrast, the data points for all but one of the radio
galaxies are above this correlation; their excess radio emission is associated
with AGN jets. Only the galaxy B2 1318+34 (aka IC 883), which is
classified optically as a LINER galaxy, has a $L_{\rm 1.4GHz}/L_{\rm
IR}$ ratio consistent with the infrared-radio correlation.  The LINER
classification is commonly an indication of ionization of an AGN with
a shallow power law, however, the spatial coincidence between the radio
(Condon et al. 1991) and CO (Downes \& Solomon 1998) emission in the disk
of IC 883 is evidence in favor of the radio emission being associated
with a starburst population. Given this, B2 1318+34 will be denoted by
a symbol different from the other radio galaxies in Figures 6 and 8.

\subsection{Global CO-Infrared Properties}

Figure 6 shows two common plots of infrared and CO luminosity for
extragalactic sources. The first plot is the CO luminosity versus the
infrared luminosity of the radio galaxies and PG QSOs, plotted with
infrared galaxies and elliptical galaxies with CO detections. In addition,
arrows have been plotted to represent 3$\sigma$ $L'_{\rm CO}$ upper limits
for the radio galaxies with CO nondetections in the present survey.  There
are two noteworthy features in Figure 6a. First, the PG QSO population
appear to have low CO luminosities for their infrared luminosity relative
to infrared galaxies (see also Evans et al. 2001). Second, by contrast,
the radio galaxies are well distributed throughout the area of the
plot occupied by infrared and elliptical galaxies, with the three newly
detected radio galaxies (3C 31, PKS 0502-10, and 3C 293) having high CO
luminosities for their infrared luminosity. These features are better
represented in Figure 6b, which is a plot of the $L_{\rm IR} / L'_{\rm
CO}$ versus the infrared luminosity. The ratio $L_{\rm IR} / L'_{\rm CO}$
is commonly referred to as the star formation efficiency, i.e., it is the
measure of energy from massive stars heating the dust normalized by the
amount of molecular gas available to form new stars. Thus, for highly
efficient starbursts (i.e., high $L_{\rm IR} / L'_{\rm CO}$), a large
number of massive stars are produced from the available gas. The QSOs
have high $L_{\rm IR} / L'_{\rm CO}$, indicating that either they are
producing massive stars with extremely high efficiency, or that there is
significant heating of the dust by the AGN (Evans et al. 2001).  Most of
the radio galaxies have $L_{\rm IR} / L'_{\rm CO}$ similar to those of
giant molecular clouds (GMCs: Sanders \& Mirabel 1996) in the Galaxy.
Exceptions to this include PKS 1345+12, which has a $L_{\rm IR} /
L'_{\rm CO}$ similar to the PG QSOs, and the three newly detected radio
galaxies, which have $L_{\rm IR} / L'_{\rm CO} (< 30$) similar to the
global value for local spiral galaxies (see also Evans et al. 1999a).

It thus appears that while some radio galaxies have significant amounts
of molecular gas, the rate of massive star formation is extremely
low. Indeed, if the assumption is made that the infrared luminosity
in 3C 31, PKS 0502-10, and 3C 293 is due entirely to star formation,
then the star formation rates are roughly a few solar masses per year.
Note also that the dust temperatures of 3C 31, PKS 0502-10, and 3C 293
are lower (30--45 K) than that of the previously detected radio galaxies
(55--90 K), also consistent with the idea that current massive star
formation rates in these three radio galaxies is low.

\subsection{Global X-ray Properties}

Given the amount of molecular gas in the radio galaxies detected in CO to
date, and the diversity in the apparent levels of active star formation
(as traced by $L_{\rm IR}$ and $L_{\rm IR} / L'_{\rm CO}$), it is worth
examining their hard (2--10 keV) X-ray properties to determine {\it
(a)} whether these radio galaxies have X-ray emission indicative of AGN,
and {\it (b)} whether any of the radio galaxies have sufficient column
densities of molecular gas to be significantly X-ray absorbed. The
$L_{\rm X}$-to-$L_{\rm FIR}$ ratio provides a straight-forward way of
addressing these issues -- the ratio is constant in starburst galaxies,
indicating that the starburst responsible for heating the dust in HII
region galaxies produces a constant fraction of massive X-ray binary pairs
(Grimm et al. 2003; Ranalli et al.  2003). `Excesses' in $L_{\rm X}$ /
$L_{\rm FIR}$ arise from the contribution of AGN to the X-ray emission,
making the ratio a powerful AGN vs. starburst diagnostic. However, in
objects that contain large columns of gas and dust, the X-ray emission
can be suppressed, lowering the observed $L_{\rm X} / L_{\rm FIR}$.

For the purposes of the present discussion, the available radio galaxy
data of our sample have been compiled from the literature and added to
the $L_{\rm X}$ vs. $L_{\rm FIR}$ plot of Trentham et al. (2005), which
is a compilation of data of local LINERs, Seyferts, HII-region
galaxies, infrared galaxies, and PG QSOs. The result is plotted in
Figure 7; Figure 7a shows the data plotted in the manner typically
done by other authors, i.e.,  $L_{\rm X}$ vs. $L_{\rm FIR}$, and
Figure 7b shows the data plotted in terms of the $L_{\rm X} / L_{\rm
FIR}$ versus $L_{\rm FIR}$. A least-squares fit has been applied to
the HII region galaxy data.  Several of the infrared galaxies, which
are rich in molecular gas, are below the fit and thus Compton thick.
In terms of the radio galaxies data, the Figures show that, {\it (a)}
the AGN dominates the X-ray emission from the radio galaxies in a manner
consistent with PG QSOs and lower luminosity Seyferts and LINERs, and {\it
(b)} the X-ray emission from the AGN does not seem to be significantly
suppressed, and thus these radio galaxies do not reside in Compton-thick
environments. Indeed, several of the radio galaxies have measured
column densities in the literature -- the derived column densities from
these X-ray observations are
$\sim$ $10^{21-22}$ cm$^{-2}$ (Leighly et al. 1997; Sambruna et al.
1999; and Odea et al. 2000).

\subsection{CO Luminosity and $L_{\rm IR} / L'_{\rm CO}$ versus Radio
Power}

Another issue relevant to the present sample of galaxies is whether the
molecular gas properties are correlated with the radio output of the
radio galaxies. I.e., if molecular gas fuels AGN activity, there may be a
correlation between $L'_{\rm CO}$ and the energy emitted by these AGN at
radio wavelengths. Figure 8a is a plot of $L'_{\rm CO}$ versus $L_{\rm
1.4GHz}$ of the sample of radio galaxies; comparative samples of infrared
galaxies, PG QSOs, and ellipticals are also plotted. The infrared galaxy
and PG QSO data are correlated, with higher $L'_{\rm CO}$ corresponding
to higher $L_{\rm 1.4GHz}$.  Such a correlation can be understood, for
example, if a significant fraction of the 1.4 GHz emission is generated
via star formation -- the higher the molecular gas content, the higher
the number of supernovae generating radio emission. The radio galaxies
occupy a different location on this plot, having high $L_{\rm 1.4GHz}$
for their $L'_{\rm CO}$ relative to infrared galaxies and QSOs. It
is also clear from the radio galaxy data that no obvious correlation
exists between $L_{\rm 1.4GHz}$ and $L'_{\rm CO}$. This result may simply
illustrate the fact that, if a correlation between the AGN radio output
and molecular gas mass exists, it may only exist for the component of
the molecular gas in the nuclear region (near the AGN), and not the
global molecular gas mass measured by these observations.

Figure 8b is a plot of $L_{\rm IR} / L'_{\rm CO}$ versus $L_{\rm
1.4GHz}$. In terms of the distribution of infrared galaxies and QSO
data, the plot is reminiscent of Figure 6b -- $L_{\rm IR} / L'_{\rm CO}$
increases with increasing $L_{\rm 1.4GHz}$, and the QSOs and warm infrared
galaxies reside primarily at high $L_{\rm IR} / L'_{\rm CO}$ and $L_{\rm
1.4GHz}$ relative to cool infrared galaxies.  The radio galaxies occupy
a different location on the plot, having high $L_{\rm 1.4GHz}$ for their
$L_{\rm IR} / L'_{\rm CO}$, but no obvious correlation among the radio
galaxy data exists.  Thus, regardless of whether stars or AGN in the
host galaxy of these radio galaxies are heating the dust, the fueling
efficiency factor $L_{\rm IR} / L'_{\rm CO}$ is not correlated with the
total radio output.

\subsection{High Mass Star Formation}

All but one of the radio galaxies detected as part of this survey have
molecular gas masses in excess of $1\times10^9$ M$_\odot$, indicating
that there are large fuel reservoirs for star formation and possibly
AGN activity. The presence of
molecular gas, in itself, is not evidence of active star formation. High
mass star formation in the Milky Way occurs in the cores of giant
molecular clouds (GMCs) where the molecular gas densities exceed $\sim
10^5$ cm$^{-3}$.

Preliminary observations have been done of PKS 0502-10 and B2 0722+30 to
provide additional information about the state of the molecular gas. The
first observations are of the CO($2\to1$) emission line; the $L'_{\rm
CO(2\to1)} / L'_{\rm CO(1\to0)}$ ratios of PKS 0502-10 and B2 0722+30
are 0.6 and 0.7, respectively.  Such low values of $L'_{\rm CO(2\to1)}
/ L'_{\rm CO(1\to0)}$ are suggestive of subthermally excited CO, 
i.e., low molecular gas temperatures and densities.

The second set of preliminary observations designed to probe the state
of the gas in these radio galaxies are of the high density
tracer HCN. HCN is seen in abundance when the molecular gas densities
are on the order of $10^5$ cm$^{-3}$, and thus is found in the cores of
GMCs where massive star formation is commencing. Gao \&
Solomon (2004) have presented an HCN survey of infrared galaxies, and have
calculated an average $L_{\rm IR} / L'_{\rm HCN}$ ratio of $\sim 970\pm600$.
The HCN($1\to0$) observations of two radio galaxies yielded only upper limits,
translating into $L_{\rm IR} / L'_{\rm HCN}$ lower limits of $\sim$
550 and 3700 for PKS 0502-10 and B2 0722+30, respectively. Thus, while
the $L_{\rm IR} / L'_{\rm HCN}$ of PKS 0502-10 does not provide a
useful constraint, the high $L_{\rm IR} / L'_{\rm HCN}$ of B2 0722-30
is suggestive of either highly efficient massive star formation or
significant heating of dust by the AGN.

\subsection{Are Fanaroff-Riley II Radio Galaxies Relatively Molecular
Gas Poor?}

One of the goals of this survey was to obtain detections of CO emission
from radio galaxies of all morphological types. This survey yielded a CO
detection in only one Fanaroff-Riley II (edge-brightened radio morphology:
Fanaroff \& Riley 1974) radio galaxy, 3C 293. However, in terms of its
radio properties, 3C 293 is not a representative F-R II. The galaxy
has a low radio power ($L_{\rm 1.4GHz} \sim 2\times10^{25}$ W m$^{-2}$
Hz$^{-1}$) relative to the bulk of F-R II radio galaxies. In addition,
3C 293 has a complex steep spectrum core which shows evidence of a
previous radio outburst (Akujor et al.  1996).

The low CO detection rate of F-R II galaxies ($\sim$ 8\%) relative to
F-R I and compact radio source galaxies ($\sim$ 35\%) may be a simple
consequence of larger distances of F-R II galaxies, on average, relative
to lower radio power radio galaxies in the sample.  This explanation
is unlikely; Figure 6a shows the $L'_{\rm CO}$ of the detected radio
galaxies and the $L'_{\rm CO}$ upper limits of the undetected galaxies.
The upper limits of all of the non-detected F-R II galaxies are below
$2.5\times10^9$ K km s$^{-1}$ pc$^2$, and three out of seven (excluding
B2 1318+34) of the CO-detected F-R I and compact radio galaxies are
above this line.  Thus, though the F-R II galaxies comprise 36\%
of the sample, there are no additional F-R II galaxies in the sample
with molecular gas masses comparable to either 3C 293, or the most
gas-rich F-R I and radio compact galaxies (PKS 0502-10, TXS 1506+345,
and PKS 1345+12) in the sample. This result, of course, relies on the
assumption that F-R II radio galaxies have CO emission line widths of
$\sim 500$ km s$^{-1}$; if the line widths are sufficiently broad, then
the limited velocity bandwidth (i.e., $\sim$ 1400 km s$^{-1}$ at 3mm)
available with single-dish telescopes such as the 12m Telescope would
make the CO detections difficult.

The present CO survey has focussed on a sample of IRAS-detected
radio galaxies using a fairly moderate size single-dish millimeter
telescope. The results presented here can be improved upon in the
following ways: {\it (i)} CO surveys of larger samples of ISO and
Spitzer Space Telescope-detected radio galaxies using larger single-dish
millimeter-wave telescopes or interferometers.  {\it (ii)} Volume-limited
CO surveys of radio galaxies (and QSOs) similar to those initiated by
Lim et al. (1999) and Scoville et al. (2003). Such surveys will provide a more
complete view of the molecular gas properties of radio (and UV-selected)
AGN.  {\it (iii)} Interferometric observations of the CO-detected radio
galaxies to measure the extent, surface density, and kinematics of the
molecular gas relative to the stellar population and the ionized/neutral
gas (Inoue et al. 1996; Evans et al. 1999a,b; Okuda et al. 2005; Evans et
al. 2005a). Finally, continued investigations of star formation in radio
galaxies using diagnostics at multiple wavelengths will provide a more
complete picture of the putative starburst-AGN connection.  Recent optical
spectroscopy of three of the galaxies in the sample (3C 293, 3C 305,
and PKS 1345+12: Tadhunter et al. 2005) have provided encouraging results.

\section{Summary}

The survey presented here is a follow-up to the Mazzarella et al. (1993)
study of IRAS-detected radio galaxies, which yielded new CO detections in
three radio galaxies. The survey has increased the number of low-redshift
radio galaxies observed in the prior survey by a factor of four.
The following conclusions have been reached:

(1) Three additional radio galaxies (3C 31, PKS 0502-10, 3C 293) have
been detected in CO($1\to0$), yielding a total of 9 detections out
of 33 radio galaxies. The CO detections translate into molecular gas
masses of $0.4-7\times10^9$ M$_\odot$, with 3$\sigma$ upper limits for
nondetections in the range of $< 4\times10^8$ M$_\odot$ for a $z \sim
0.02$ galaxy to $<3.3\times10^9$ M$_\odot$ for a $z \sim 0.1$ galaxy
($\alpha$ = 1.5 $M_{\odot}$ [K km s$^{-1}$ pc$^2]^{-1}$. Thus,
the upper limits of the more distant members of
the sample do not rule out molecular gas mass as high as that of the Milky Way 
($\sim 3\times10^9$ M$_\odot$).

(2) Many of the CO lines have double-peaked profiles, which is indicative
of either CO absorption towards the AGN, or CO distributed in a molecular
ring.

(3) The CO line widths are extremely broad. The average velocity dispersion
of the CO-detected radio galaxies is $\Delta v_{\rm FWHM} = 500\pm130$
km s$^{-1}$, which exceeds the average line width of both ultraluminous
infrared galaxies (= $300\pm90$ km s$^{-1}$) and PG QSOs (= $260\pm160$ km
s$^{-1}$). This is an indication that the CO-luminous radio galaxies
are massive.

(4) With the exception of the edge-on spiral galaxy B2 0722+30, optical
images show the molecular gas-rich radio galaxies to have close companions
and/or tidal features consistent with those of infrared-luminous
interacting and merging galaxy systems.

(5) Most of the radio galaxies have $L_{\rm IR} / L'_{\rm CO}$ similar
to GMCs.  Exceptions to this include {\it (i)} PKS 1345+12, which has a
$L_{\rm IR} / L'_{\rm CO} (\sim 370$) comparable to the PG QSOs, 
likely an indication of highly efficient massive star production
or of significant dust heating by the AGN, and {(\it ii)} 3C 31, PKS
0502-10, and 3C 293, which have low $L_{\rm IR} / L'_{\rm CO}$ ($<$ 30)
consistent with the global $L_{\rm IR} / L'_{\rm CO}$ of spiral galaxies.

(6) All of the radio galaxies have measured values of $L_{\rm X}/L_{\rm FIR}$
significantly higher than the average for HII-region like galaxies. Thus,
the AGN is the major contributor to the X-ray emission, and the AGN
do not reside in Compton-thick environments.

(7) No correlation exists between $L'_{\rm CO}$ or $L_{\rm IR} / L'_{\rm
CO}$ and $L_{\rm 1.4GHz}$ of the radio galaxies.

(8) Only one F-R II radio galaxy (3C 293) was detected, yielding an
8\% CO detection rate for F-R II galaxies. By
comparison, $\sim 35$\% of the F-R I and compact radio source galaxies were
detected. The $L'_{\rm CO}$ upper limits of the undetected F-R II galaxies
are sufficiently small to support the conclusion that either low-redshift
F-R II galaxies are gas-poor compared with F-R I and radio compact
galaxies, or that the CO line widths of the undetected F-R II galaxies are
broad enough to make their detection difficult with a single-dish
telescope.

\acknowledgements

We thank the telescope operators and staff of the NRAO 12m and IRAM
30 telescopes for their support both during and after the observations
were obtained, and the anonymous referee for many useful comments and
suggestions. ASE also thanks N. Trentham for useful discussions and
assistance. ASE was supported by NSF grant AST 02-06262 and the 2002
NASA/ASEE Faculty Fellowship. DTF, JMM, and JAS were supported by the
Jet Propulsion Laboratory, California Institute of Technology, under
contract with NASA. D.B.S. gratefully acknowledges the hospitality of
the Max-Plank Institut fur Extraterrestrische Physik and the Alexander
von Humboldt Stiftung for a Humboldt senior award, and partial financial
support from NASA grant GO-8190.01-97A.  This research has made use of
the NASA/IPAC Extragalactic Database (NED) which is operated by the Jet
Propulsion Laboratory, California Institute of Technology, under contract
with the National Aeronautics and Space Administration.

\vfill\eject

\centerline{Figure Captions}

\vskip 0.3in

\noindent
Figure 1. (a) NRAO 12m and IRAM 30m CO(1$\to$0) spectrum of 31 of the
33 radio galaxies in the survey. The intensity scale is in units of
main beam brightness temperature. A linear baseline has been subtracted
from each spectrum; the baseline subtraction is performed outside of
the velocity range of emission lines when they are present.  With the
exception of PKS 0502-10, the zero velocity corresponds to the redshift
of the radio galaxy listed in Table 1. The zero velocity redshift of PKS
0502-10 is 0.0396. (b) IRAM 30m CO(1$\to0$) spectrum of PKS 1345+12. The
spectrum confirm the general features of the 12m spectrum of Figure 1a,
but with significantly higher signal-to-noise.

\noindent
Figure 2. JCMT 15m and IRAM 30m telescope CO($2\to1$) spectra of PKS
0502-10, B2 0722+30, and Cygnus A. The intensity scale is in units of main
beam brightness temperature (IRAM) and antenna temperature corrected for
aperture losses (JCMT). A linear baseline has been subtracted from each
spectrum; the baseline subtraction is performed outside of the velocity
range of emission lines when they are present.  Zero velocity corresponds
to the redshift of the radio galaxy listed in Table 3.

\noindent
Figure 3. IRAM 30m HCN($1\to0$) spectra of PKS 0502-10 and B2 0722+30.
The intensity scale is in units of main beam brightness temperature. A
linear baseline has been subtracted from each spectrum; the baseline
subtraction is performed outside of the velocity range where the HCN
emission line is expected to be.  Zero velocity corresponds to the
redshift of the radio galaxy listed in Table 3.

\noindent
Figure 4. False-color {\it U$^{\prime}$}, {\it B}, and {\it I}-band
images of seven of the radio galaxies detected in CO($1\to0$). The
{\it U}, {\it B}, and {\it I}-band data are displayed as blue, green,
and red, respectively.  North is up, and east is to the left.

\noindent
Figure 5. The ratio of 1.4 GHz to infrared luminosity versus the infrared
luminosity of infrared luminous radio galaxies, a sample of infrared
galaxies, PG QSOs and elliptical galaxies detected in CO. References for
the infrared data are as follows: infrared galaxies, Mazzarella et al.
(1993); PG QSOs, Sanders et al. (1989) and Haas et al. (2003); elliptical
galaxies (Moshir et al. 1990). References for the 1.4 GHz data are as
follows: infrared galaxies, Condon et al. (1990, 1991, 1992); elliptical
galaxies (White \& Becker 1992; Becker et al. 1995; Condon et al. 1998);
radio galaxies (White \& Becker 1992; Becker et al. 1995); PG QSOs
(White \& Becker 1992; Barvainis, Lonsdale, \& Antonucci 1996; Condon
et al. 1998).

\noindent
Figure 6. (a) CO vs. infrared luminosity plot of infrared luminous radio
galaxies, a sample of infrared galaxies, PG QSOs and elliptical galaxies
detected in CO.  References for CO data are as follows: infrared galaxies,
Mazzarella et al. (1993); PG QSOs, Sanders et al. (1988c), Barvainis et
al. (1989), Evans et al. (2001), and Scoville et al. (2003); elliptical
galaxies: Wiklind \& Combes (1997).  (b) the ratio of infrared to CO
luminosity vs. infrared luminosity for the same sample of galaxies as
in 6(a). In both Figures, the galaxy B2 1318+34 is denoted by an unfilled
triangle.

\noindent
Figure 7. (a) Hard (2-10 keV) X-ray luminosity vs. far-infrared luminosity
plot of the infrared luminous radio galaxies and a comparison sample
of infrared galaxies, PG QSOs, and less luminous galaxies classified
as HII region-like galaxies, Seyferts, and LINERs.  The far-infrared
luminosities have been calculated using data in NED. The radio galaxies
in the Figure are NGC 315, 3C 84, 3C 111, 3C 120, PKS 0634-20,
Hydra A, M87, PKS 1345+12, 3C 293, 3C 321, NGC 6251, 3C 390.3, and
3C 405. References for the X-ray data are as follows: HII, Seyferts,
and LINERs, Terashima et al. (2002) and Grimm et al. (2003); infrared
luminous galaxies, Iwasawa \& Comastri (1998), Boller (2002), Dalla Ceca
(2002), Franceschini et al. (2003), and Sanders et al.  (2005); radio
galaxies, Churazov et al. (2003); Di Matteo et al. (2003); Leighly et
al. (1997); McNamara et al. (2000); Odea et al. (2000); and Sambruna et
al. (1999). The X-ray luminosity of 3C 293 ($L_X \sim 3\times10^{42}$
erg s$^{-1}$) was calculated by us from archival ASCA data. (b) Ratio of
the X-ray to far-infrared luminosity versus the far-infrared luminosity
for the same sample of galaxies as in Figure 7(a). In both (a) and (b),
radio galaxies and PG QSOs with CO detections are encircled. Adapted
from Trentham et al. (2005).

\noindent
Figure 8. (a) The CO luminosity vs. the radio power of the infrared
luminous radio galaxies and a comparison sample of infrared galaxies,
PG QSOs, and elliptical galaxies.  (b) $L_{\rm IR} / L'_{\rm CO}$
vs. the radio power for the same sample of galaxies plotted in 8a. In
both Figures, the galaxy B2 1318+34 is denoted by an unfilled triangle.

\begin{deluxetable}{lrcccrrllc}
\pagestyle{empty}
\scriptsize
\tablenum{1}
\tablewidth{0pt}
\tablecaption{Source List}
\tablehead{
\multicolumn{1}{c}{} &
\multicolumn{2}{c}{Coordinates (J2000.0)} &
\multicolumn{1}{c}{} &
\multicolumn{1}{c}{} &
\multicolumn{1}{c}{$f_{60\mu{\rm m}}$} &
\multicolumn{1}{c}{$f_{100\mu{\rm m}}$} &
\multicolumn{1}{c}{$L_{\rm IR}$$^{\rm a}$} &
\multicolumn{1}{c}{$L_{\rm FIR}$$^{\rm a}$} &
\multicolumn{1}{c}{$P_{1.5{\rm GHz}}$}  \nl
\multicolumn{1}{c}{Source} &
\multicolumn{1}{c}{RA} &
\multicolumn{1}{c}{Dec} &
\multicolumn{1}{c}{z} &
\multicolumn{1}{c}{FR} &
\multicolumn{1}{c}{(mJy)} &
\multicolumn{1}{c}{(mJy)} &
\multicolumn{1}{c}{($L_{_\odot}$)} &
\multicolumn{1}{c}{($L_{_\odot}$)} &
\multicolumn{1}{c}{(W m$^{-2}$ Hz$^{-1}$)}}
\startdata
NGC 315 & 00h57m48.85s &  +30d21m08.63s & 0.0165 & I & 368 & 460 & 7.2$\times10^{9}$ & 2.3$\times10^{9}$ & 1.6$\times10^{24}$ \nl
3C 31 & 01h07m24.98s &  +32d24m44.77s & 0.0167 & I & 435 & 1675 & 7.4$\times10^{9}$ & 4.7$\times10^{9}$ & 2.5$\times10^{24}$ \nl
4C 31.04 & 01h19m35.00s  &  +32d10m50.03s & 0.0590 & C & 150 & 524 & 3.7$\times10^{10}$ & 1.9$\times10^{10}$ & 1.7$\times10^{25}$ \nl
NGC 741 & 01h56m21.05s &  +05d37m44.17s & 0.0185 & C & 214 & 777 & 7.0$\times10^{9}$ & 2.7$\times10^{9}$ & 6.0$\times10^{23}$ \nl
3C 84 & 03h19m48.09s &  +41d30m42.47s & 0.0176 & I & 7427 & 8267 & 1.6$\times10^{11}$ & 5.1$\times10^{10}$ & 7.3$\times10^{24}$ \nl
3C 88 & 03h27m54.17s &  +02d33m41.85s & 0.0302 & II & 180 & 816 & 1.7$\times10^{10}$ & 7.0$\times10^{9}$ & 8.5$\times10^{24}$ \nl
3C 111 & 04h18m21.07s &  +38d01m32.59s & 0.0485 & II & 321 & $<$2250 & 9.0$\times10^{10}$ & 2.8$\times10^{10}$ & 6.5$\times10^{25}$ \nl
3C 120 & 04h33m11.09s &  +05d21m16.03s & 0.0331 & I & 1383 & 1937 & 1.2$\times10^{11}$ & 3.6$\times10^{10}$ & 9.0$\times10^{24}$ \nl
PKS 0502-10 & 05h04m52.99s &  -10d14m52.18s & 0.0410 & ? & 688 & 1370 & 6.1$\times10^{10}$ & 3.0$\times10^{10}$ & 4.6$\times10^{24}$ \nl
PKS 0634-20 & 06h36m32.24s &  -20d34m52.92s & 0.0544 & II & 484 & 1150 & 9.5$\times10^{10}$ & 4.3$\times10^{10}$ & 3.3$\times10^{25}$ \nl
B2 0648+27 & 06h42m02.82s &  +27d28m21.69s & 0.0414 & C & 2633 & 1529 & 2.2$\times10^{11}$ & 8.7$\times10^{10}$ & 4.9$\times10^{23}$ \nl
B2 0722+30 & 07h25m37.35s &  +29d57m14.71s & 0.0188 & I & 3108 & 4999 & 5.2$\times10^{10}$ & 2.8$\times10^{10}$ & 8.2$\times10^{22}$ \nl
4C 29.30 & 08h40m02.44s &  +29d49m02.87s & 0.0650 & I & 472 & 595 & 1.1$\times10^{11}$ & 4.7$\times10^{10}$ & 3.3$\times10^{23}$ \nl
Hydra A & 09h18m05.69s &  -12d05m45.45s & 0.0538 & I & 155 & 416 & 2.8$\times10^{10}$ & 1.4$\times10^{10}$ & 2.5$\times10^{26}$ \nl
3C272.1 & 12h25m03.78s &  +12d53m13.1s & 0.0035 & I & 556 & 1024 & 5.6$\times10^{8}$ & 1.8$\times10^{8}$ & 1.0$\times10^{23}$ \nl
M 87 & 12h30m49.35s &  +12d23m28.06s & 0.0047 & I & 546 & 559 & 1.1$\times10^{9}$ & 2.6$\times10^{8}$ & 7.2$\times10^{24}$ \nl
B2 1318+34 & 13h20m35.20s &  +34d08m13.56s & 0.0233 & I & 16070 & 20682 & 3.4$\times10^{11}$ & 2.0$\times10^{11}$ & 8.9$\times10^{22}$ \nl
3C 285 & 13h21m17.78s  & +42d35m14.23s & 0.0789 & II & 253 & 637 & 1.8$\times10^{11}$ & 5.0$\times10^{10}$ & 2.6$\times10^{25}$ \nl
PKS 1345+12 & 13h47m33.38s &  +12d17m24.18s & 0.1224 & C & 2098 & 1738 & 1.5$\times10^{12}$ & 6.7$\times10^{11}$ & 1.5$\times10^{26}$ \nl
3C 293 & 13h52m17.86s &  +31d26m46.67s & 0.0450 & II & 233 & 621 & 2.9$\times10^{10}$ & 1.5$\times10^{10}$ & 1.7$\times10^{25}$ \nl
OQ 208 & 14h07m00.39s &  +28d27m14.73s & 0.0768 & C & 753 & 1029 & 4.1$\times10^{11}$ & 1.1$\times10^{11}$ & 8.3$\times10^{24}$ \nl
3C 305 & 14h49m21.46s  & +63d16m16.14s & 0.0410 & I & 298 & 450 & 2.4$\times10^{10}$ & 1.2$\times10^{10}$ & 9.4$\times10^{24}$ \nl
TXS 1506+345 & 15h08m05.64s &  +34d23m22.75s & 0.0449 & I & 2647 & 3956 & 2.3$\times10^{11}$ & 1.3$\times10^{11}$ & 4.5$\times10^{23}$ \nl
PKS B1518+045 & 15h21m22.61s  & +04d20m29.21s & 0.0513  & ? & $<$120 & 968 & 4.6$\times10^{10}$ & 1.8$\times10^{10}$ & 2.1$\times10^{25}$ \nl
3C 321 & 15h31m43.45s &  +24d04m18.70s & 0.0960 & II & 1067 & 961 & 5.0$\times10^{11}$ & 2.1$\times10^{11}$ & 6.2$\times10^{25}$ \nl
3C 327 & 16h02m27.39s  & +01d57m55.70s & 0.1048 & II & 670 & 371 & 4.1$\times10^{11}$ & 1.4$\times10^{11}$ & 1.9$\times10^{26}$ \nl
NGC 6251 & 16h32m31.97s   &   +82d32m16.46s & 0.0249 & I & 188 & 600 & 5.7$\times10^{9}$ & 2.7$\times10^{9}$ & 2.6$\times10^{24}$ \nl
B2 1707+34 & 17h09m38.37s &  +34d25m53.76s & 0.0808 & II & 464 & 841 & 2.0$\times10^{11}$ & 8.2$\times10^{10}$ & 8.4$\times10^{24}$ \nl
3C 390.3 & 18h42m08.92s &  +79d46m17.27s & 0.0561 & II & 250 & 340 & 1.3$\times10^{11}$ & 1.9$\times10^{10}$ & 7.2$\times10^{25}$ \nl
3C 402 & 19h41m46.00s &  +50d35m44.86s & 0.0254 & II & 257 & 1052 & 1.1$\times10^{10}$ & 6.6$\times10^{9}$ & 3.1$\times10^{24}$ \nl
3C 403 & 19h52m15.36s &  +02d30m28.14s & 0.0590 & II & 441 & $<$500 & 1.5$\times10^{11}$ & 3.0$\times10^{10}$ & 3.9$\times10^{25}$ \nl
3C 405 & 19h59m28.51s  & +40d44m01.70s & 0.0556 & II & 2847 & $<$1800 & 4.0$\times10^{11}$ & 1.6$\times10^{11}$ & 9.4$\times10^{27}$ \nl
3C 433 & 21h23m44.53s  & +25d04m09.99s & 0.1016 & I & 299 & $<$1200 & 2.9$\times10^{11}$ & 8.8$\times10^{10}$ & 2.3$\times10^{24}$ \nl
\enddata
\tablenotetext{a}{Both $L_{\rm IR}$ and $L_{\rm FIR}$ are calculated
using the prescription in Sanders \& Mirabel (1996). In cases where one
or more IRAS flux densities have only upper limit values, $L_{\rm IR}$
and $L_{\rm FIR}$ are calculated by setting the flux density value equal
to the upper limit value and by setting the flux density value equal to
zero, then taking the average of these two determinations.}
\tablerefs{Fanaroff-Riley classications: Golombek et al. 1988;  Baum, Heckman,
\& van Breugel 1992;
http://www.jb.man.ac.uk/atlas/.
Flux Densities: Golombek et al. 1988; Redshifts: Golombek et al. 1988; Mirabel et al. 1989; Mazzarella et al. 1993.}
\end{deluxetable}

\begin{deluxetable}{lrlcrc}
\pagestyle{empty}
\tablenum{2}
\tablewidth{0pt}
\tablecaption{Journal of Observations}
\tablehead{
\multicolumn{1}{c}{} &
\multicolumn{1}{c}{} &
\multicolumn{1}{c}{} &
\multicolumn{1}{c}{Date} &
\multicolumn{1}{c}{Time} &
\multicolumn{1}{c}{T$_{sys}$}  \nl
\multicolumn{1}{c}{Source} &
\multicolumn{1}{c}{Site} &
\multicolumn{1}{c}{Transition} &
\multicolumn{1}{c}{(mmm-yy)} &
\multicolumn{1}{c}{(hr)} &
\multicolumn{1}{c}{(K)}}
\startdata
NGC 315	     &  12m &	CO(1$\to0$) &	Oct-97 & 3.0 & 	420 \nl	   
3C 31	     &  12m &	CO(1$\to0$) &	Jan-96 & 17.4 & 	272   \nl
	     &  12m &	CO(1$\to0$) &	Jun-97 & 18.4 & 	333   \nl
	     &  12m &	CO(1$\to0$) &	Oct-97 & 4.6 & 	273   \nl
4C 31.04 &      12m &	CO(1$\to0$) &	Jan-96 & 29.6 & 	199   \nl
NGC 741	&       12m &	CO(1$\to0$) &	Jun-97 & 12.4 & 	371   \nl
3C 84	&       12m &	CO(1$\to0$) &	Jun-97 & 4.8 & 	318   \nl
3C 88	&       12m &	CO(1$\to0$) &	Oct-97 & 7.2 & 	248   \nl
3C 111	&       12m &	CO(1$\to0$) &	Sep-95 & 10.2 & 	260   \nl
3C 120	&       12m &	CO(1$\to0$) &	Oct-97 & 19.6 & 	259   \nl
PKS 0502-10 &   12m &	CO(1$\to0$) &	Sep-95 & 14.8 & 	369   \nl
	   &    12m &	CO(1$\to0$) &	Jan-96 & 9.4 & 	256   \nl
	   &    15m &	CO(2$\to1$) &	Sep-95 & 4.0 & 	411   \nl
           &    30m &   HCN(1$\to0$) &  Dec-03 & 1.8 &  120     \nl
PKS 0634-20 &   12m &	CO(1$\to0$) &	Jun-97 & 5.2 & 	372   \nl
	   &    12m &	CO(1$\to0$) &	Oct-97 & 12.6 & 	256   \nl
B2 0648+27 &    12m &	CO(1$\to0$) &	Sep-95 & 24.6 & 	319   \nl
	  &     12m &	CO(1$\to0$) &	Jan-96 & 2.6 & 	237   \nl
B2 0722+30 &    12m &	CO(1$\to0$) &	Oct-97 & 6.6 & 	299   \nl
	  &     15m &	CO(2$\to1$) &	Sep-95 & 1.8 & 	469   \nl
          &     30m &   HCN(1$\to0$) &  Dec-03 & 1.8  & 83       \nl
4C 29.30 &      12m &	CO(1$\to0$) &	Jan-96 & 12.2 & 	220   \nl
	&       12m &	CO(1$\to0$) &	Jun-97 & 30.8 & 	232   \nl
	&       12m &	CO(1$\to0$) &	Oct-97 & 8.2 & 	196   \nl
Hydra A	&       12m &	CO(1$\to0$) &	Jun-97 & 5.2 & 	288   \nl
	&       12m &	CO(1$\to0$) &	Oct-97 & 9.4 & 	224   \nl
M 87	&       12m &	CO(1$\to0$) &	Dec-92 & 21.9 & 	368   \nl
B2 1318+34 &   12m &	CO(1$\to0$) &	Oct-97 & 1.8 & 	256   \nl
3C 285	   &    12m &	CO(1$\to0$) &	Jan-96 & 49.6 & 	210   \nl
PKS 1345+12 &   12m &	CO(1$\to0$) &	Jun-97 & 30.1 & 	256   \nl
	   &    12m &	CO(1$\to0$) &	Jun-00 & 4.3 & 	200   \nl
3C 293	   &    12m &	CO(1$\to0$) &	Jan-96 & 17.6 & 	240   \nl
OQ 208	   &    12m &	CO(1$\to0$) &	Sep-95 & 12.8 & 	261   \nl
3C 305	   &    12m &	CO(1$\to0$) &	Oct-97 & 16.0 & 	261   \nl
TXS 1506+345 &   12m &	CO(1$\to0$) &	Oct-97 & 8.4 & 	211   \nl
PKS B1518+045 &   12m &	CO(1$\to0$) &	Jun-97 & 9.4 & 	269   \nl
3C 321	   &    12m &	CO(1$\to0$) &	Sep-95 & 23.6 & 	276   \nl
	   &    12m &	CO(1$\to0$) &	Jun-97 & 16.4 & 	249   \nl
3C 327	   &    12m &	CO(1$\to0$) &	Oct-97 & 5.6 & 	321   \nl
	   &    12m &	CO(1$\to0$) &	Jun-00 & 16.2 & 	205   \nl
B2 1707+34 &   12m &	CO(1$\to0$) &	Jun-97 & 13.0 & 	252   \nl
	   &    12m &	CO(1$\to0$) &	Oct-97 & 12.0 & 	208   \nl
3C 390.3 &      12m &	CO(1$\to0$) &	Jun-97 & 16.4 & 	356   \nl
	&       12m &	CO(1$\to0$) &	Oct-97 & 5.2 & 	464   \nl
3C 402	&       12m &	CO(1$\to0$) &	Oct-97 & 6.0 & 	248   \nl
3C 403	&       12m &	CO(1$\to0$) &	Jun-97 & 11.4 & 	254   \nl
3C 405	&       30m &	CO(1$\to0$) &	Nov-95 & 5.9 & 	186   \nl
	&       30m &	CO(2$\to1$) &   Nov-95 & 3.3 & 	321   \nl
3C 433	&       12m &	CO(1$\to0$) &	Jan-96 & 11.6 & 	200   \nl
	&       12m &	CO(1$\to0$) &	Oct-97 & 13.0 & 	179   \nl
\enddata
\end{deluxetable}

\begin{deluxetable}{lrlllcrrrr}
\tablenum{3}
\scriptsize
\tablewidth{0pt}
\tablecaption{Molecular Gas Emission Line Properties}
\tablehead{
\multicolumn{1}{c}{Source} &
\multicolumn{1}{c}{$D_{\rm L}^{\rm ~~~a}$} &
\multicolumn{1}{c}{line} &
\multicolumn{1}{c}{Site} &
\multicolumn{1}{c}{$z_{\rm CO}$} &
\multicolumn{1}{c}{$\Delta v_{\rm FWHM}$} &
\multicolumn{1}{c}{$T_{\rm mb} \Delta v^{\rm ~b}$} &
\multicolumn{1}{c}{$S_{\rm line} \Delta v$} &
\multicolumn{1}{c}{${L'_{\rm line}}$} &
\multicolumn{1}{c}{$M$(H$_2)^{\rm ~c}$}\nl
\multicolumn{1}{c}{} &
\multicolumn{1}{c}{(Mpc)} &
\multicolumn{1}{c}{} &
\multicolumn{1}{c}{} &
\multicolumn{1}{c}{} &
\multicolumn{1}{c}{(km s$^{-1}$)} &
\multicolumn{1}{c}{(K km s$^{-1}$)} &
\multicolumn{1}{c}{(Jy km s$^{-1}$)} &
\multicolumn{1}{c}{(K
km s$^{-1}$ pc$^2$)} &
\multicolumn{1}{c}{($M_\odot$)}
}
\startdata

NGC 315 & 66 & CO(1$\to$0) & 12m & \nodata & \nodata & $<$0.91  & $<$23.0  & $<$2.4$\times10^{8}$  & $<$3.6$\times10^{8}$  \nl

3C 31 & 67 &  CO(1$\to$0) & 12m & 0.0169 & 450 &
1.1$\pm$0.08 & 27$\pm$2 &
2.9$\times10^{8}$ & 4.3$\times10^{8}$ \nl

4C 31.04 & 239  & CO(1$\to$0) & 12m & \nodata & \nodata & $<$0.21   & $<$5.2  & $<$6.7$\times10^{8}$  & $<$1.0$\times10^{9}$  \nl

NGC 741 & 74  & CO(1$\to$0) & 12m & \nodata & \nodata & $<$0.39   & $<$9.8  & $<$1.3$\times10^{8}$  & $<$1.9$\times10^{8}$  \nl

3C 84 & 71 &  CO(1$\to$0) & 12m & 0.0176 & 200 &
4.1$\pm$0.05 & 104$\pm$1 &
1.2$\times10^{9}$ & 1.8$\times10^{9}$\nl

3C 88 & 122 &  CO(1$\to$0) & 12m & \nodata & \nodata & $<$0.41   & $<$10.4  & $<$3.6$\times10^{8}$  & $<$5.4$\times10^{8}$  \nl

3C 111 & 196 &  CO(1$\to$0) & 12m & \nodata & \nodata & $<$0.34   & $<$8.5  & $<$7.5$\times10^{8}$  & $<$1.1$\times10^{9}$  \nl

3C 120 & 133 &  CO(1$\to$0) & 12m & 0.0331 & 550 &
1.1$\pm$0.1 & 28$\pm$2 &
1.2$\times10^{9}$ & 1.7$\times10^{9}$ \nl

PKS 0502-10 & 160  & CO(1$\to$0) &  12m & 0.0398 & 550
& 1.8$\pm$0.09 & 45$\pm$2 &
2.7$\times10^{9}$ & 4.0$\times10^{9}$ \nl

\nodata & \nodata & CO(2$\to$1) & 15m & & 550 &
4.7$\pm$0.3 & 110$\pm$7 &
1.6$\times10^{9}$ & \nodata \nl

\nodata & \nodata  & HCN(1$\to$0) & 30m & & 550 & $<$0.33   & $<$2.0  & $<$1.1$\times10^{8}$  & \nodata  \nl

PKS 0634-20 & 220  & CO(1$\to$0) & 12m & \nodata & \nodata & $<$0.38   & $<$9.6  & $<$1.0$\times10^{9}$  & $<$1.6$\times10^{9}$  \nl

B2 0648+27 & 167  & CO(1$\to$0) & 12m & \nodata & \nodata & $<$0.26   & $<$6.6  & $<$4.3$\times10^{8}$  & $<$6.4$\times10^{8}$  \nl

B2 0722+30 & 76  & CO(1$\to$0) & 12m & 0.0188 & 450
& 2.3$\pm$0.2 & 58$\pm$4 &
7.9$\times10^{8}$ & 1.2$\times10^{9}$ \nl

\nodata & \nodata  & CO(2$\to$1) & 15m & & 450 &
6.8$\pm$0.2 & 160$\pm$5 & 5.4$\times10^{8}$ & \nodata  \nl

\nodata & \nodata & HCN(1$\to$0) & 30m & & 450 & $<$0.18   & $<$1.1  & $<$1.4$\times10^{7}$  & \nodata  \nl

4C 29.30 & 264  & CO(1$\to$0) &  12m & \nodata &\nodata & $<$0.18 &   $<$4.5  & $<$7.1$\times10^{8}$  & $<$1.1$\times10^{9}$  \nl

Hydra A & 218  & CO(1$\to$0) & 12m & \nodata & \nodata & $<$0.25 &   $<$6.4  & $<$6.9$\times10^{8}$  & $<$1.0$\times10^{9}$  \nl

M 87 & 19 &  CO(1$\to$0) & 12m & \nodata & \nodata & $<$0.46 &   $<$11.7  & $<$1.0$\times10^{7}$  & $<$1.5$\times10^{7}$  \nl

B2 1318+34 & 94 & CO(1$\to$0) & 12m & 0.0233 & 400
& 7.0$\pm$0.3 & 177$\pm$7 &
3.6$\times10^{9}$ & 5.5$\times10^{9}$ \nl

3C 285 & 322 & CO(1$\to$0) & 12m & \nodata & \nodata & $<$0.20   & $<$4.9  & $<$1.1$\times10^{9}$  & $<$1.7$\times10^{9}$  \nl

PKS 1345+12 & 503 & CO(1$\to$0) & 12m & 0.1220 & 250
& 0.24$\pm$0.06 & 6.0$\pm$1.6 &
3.3$\times10^{9}$ & 4.9$\times10^{9}$ \nl

PKS 1345+12 & 503  & CO(1$\to$0) & 30m & & 400
& 1.6$\pm$0.08 & 8.0$\pm$0.4 &
4.3$\times10^{9}$ & 6.5$\times10^{9}$ \nl

3C 293 & 182 &  CO(1$\to$0) & 12m & 0.0448 & 700 &
2.2$\pm$0.09 & 56$\pm$2 &
4.3$\times10^{9}$ & 6.4$\times10^{9}$ \nl

OQ 208 & 313 & CO(1$\to$0) & 12m & \nodata & \nodata & $<$0.31   & $<$7.9  & $<$1.7$\times10^{9}$  & $<$2.6$\times10^{9}$  \nl

3C 305 & 166 &  CO(1$\to$0) & 12m & \nodata & \nodata & $<$0.34   & $<$8.6  & $<$5.4$\times10^{8}$  & $<$8.2$\times10^{8}$  \nl

TXS 1506+345 & 181  & CO(1$\to$0) & 12m & 0.0449 & 550
& 2.5$\pm$0.09 & 63$\pm$2 &
4.8$\times10^{9}$ & 7.1$\times10^{9}$ \nl

PKS B1518+045 & 208  & CO(1$\to$0) & 12m & \nodata & \nodata & $<$0.45   & $<$11.3  & $<$1.1$\times10^{9}$  & $<$1.7$\times10^{9}$  \nl

3C 321 & 393  & CO(1$\to$0) & 12m & \nodata & \nodata & $<$0.19 &   $<$4.7  & $<$1.6$\times10^{9}$  & $<$2.4$\times10^{9}$  \nl

3C 327 & 429  & CO(1$\to$0) & 12m & \nodata & \nodata & $<$0.22 &   $<$5.4  & $<$2.2$\times10^{9}$  & $<$3.3$\times10^{9}$  \nl

B2 1707+34 & 329  & CO(1$\to$0) & 12m & \nodata & \nodata & $<$0.19   & $<$4.9  & $<$1.2$\times10^{9}$  & $<$1.8$\times10^{9}$  \nl

3C 390.3 & 227  & CO(1$\to$0) & 12m & \nodata & \nodata & $<$0.41 &   $<$10.3  & $<$1.2$\times10^{9}$  & $<$1.8$\times10^{9}$  \nl

3C 402 & 102  & CO(1$\to$0) & 12m & \nodata & \nodata & $<$0.41 &   $<$10.3  & $<$2.5$\times10^{8}$  & $<$3.8$\times10^{8}$  \nl

3C 403 & 239  & CO(1$\to$0) & 12m & \nodata & \nodata & $<$0.38 &   $<$9.5  & $<$1.2$\times10^{9}$  & $<$1.9$\times10^{9}$  \nl

3C 405 & 225  & CO(1$\to$0) & 30m & \nodata & \nodata & $<$0.37 &   $<$1.9  & $<$2.1$\times10^{8}$  & $<$3.2$\times10^{8}$  \nl

3C 433 & 416 & CO(1$\to$0) & 12m & \nodata & \nodata & $<$0.16 &   $<$3.9  & $<$1.5$\times10^{9}$  & $<$2.2$\times10^{9}$  \nl

\enddata
\tablenotetext{a}{Luminosity Distance,
calculated assuming $H_0 = 75$ km s$^{-1}$ Mpc$^{-1}$,
$q_0 = 0.5$, and $\Lambda = 0$.}
\tablenotetext{b}{For all radio galaxies with CO(1$\to$0) non-detections,
$T_{\rm mb} \Delta v$ = $3 T_{\rm mb} ({\rm rms}) \Delta
v_{\rm FWHM}$, where $T_{\rm mb} ({\rm rms})$ is the root-mean-squared
main-beam temperature and $\Delta v_{\rm FWHM} = 500$ km s$^{-1}$.}
\tablenotetext{c}{Calculated assuming $\alpha = 1.5
M_{\odot}$ [K km s$^{-1}$ pc$^2]^{-1}$.}
\end{deluxetable}

\begin{deluxetable}{lllrrlcc}
\pagestyle{empty}
\scriptsize
\tablenum{4}
\tablewidth{0pt}
\tablecaption{Estimation of $M_{\rm dust}$, $M_{\rm dyn}$, and Column Densities}
\tablehead{
\multicolumn{1}{c}{} &
\multicolumn{1}{c}{$T_{\rm dust}$} &
\multicolumn{1}{c}{$M_{\rm dust}$} &
\multicolumn{1}{c}{$M({\rm H}_2) / M_{\rm dust}$} &
\multicolumn{1}{c}{$R_{\rm CO}$} &
\multicolumn{1}{c}{$M_{\rm dyn}$} &
\multicolumn{1}{c}{$M({\rm H}_2) / M_{\rm dyn}$} &
\multicolumn{1}{c}{Column Density} \nl
\multicolumn{1}{c}{Source} &
\multicolumn{1}{c}{(K)} &
\multicolumn{1}{c}{(M$_\odot$)} &
\multicolumn{1}{c}{} &
\multicolumn{1}{c}{(pc)} &
\multicolumn{1}{c}{(M$_\odot$)} &
\multicolumn{1}{c}{} &
\multicolumn{1}{c}{(cm$^{-2}$)} }
\startdata
3C 31	      &  33 & $2.7\times10^6$ & 160 &	75 &	$4.2\times10^9$ & 0.10 & $1.6\times10^{24}$ \nl
3C 84	      &  64 & $1.7\times10^6$ & 1084 &	154 &	$2.2\times10^9$ & 0.83 & $1.6\times10^{24}$ \nl
3C 120	      &  55 & $2.0\times10^6$ & 884 &	112 &	$7.6\times10^9$ & 0.23 & $3.0\times10^{24}$ \nl
PKS 0502-10 &	46 & $3.8\times10^6$ & 1075 &	185 &	$1.3\times10^{10}$ & 0.32 & $2.5\times10^{24}$ \nl
B2 0722+30 &	50 & $2.3\times10^6$ & 525 &	105 &	$4.8\times10^9$ & 0.24 & $2.2\times10^{24}$ \nl
B2 1318+34 &	58 & $9.4\times10^6$ & 572 &	222 &	$8.0\times10^9$ & 0.68 & $2.3\times10^{24}$ \nl
PKS 1345+12 &	91 & $8.1\times10^6$ & 802 &	194 &	$7.0\times10^9$ & 0.94 & $3.6\times10^{24}$ \nl
3C 293        & 40   & $3.6\times10^6$ & 1807 &  222 &	$2.5\times10^{10}$ & 0.26 & $2.8\times10^{24}$ \nl
TXS 1506+345 &	54 & $8.4\times10^6$ & 860 &	227 &	$1.6\times10^{10}$ & 0.46 & $2.9\times10^{24}$ \nl
Average     & \nodata  &   \nodata & $860\pm460$ & \nodata & \nodata & $0.45\pm0.29$ &  \nodata \nl
\enddata
\end{deluxetable}

\begin{figure}[h]
\plotfiddle{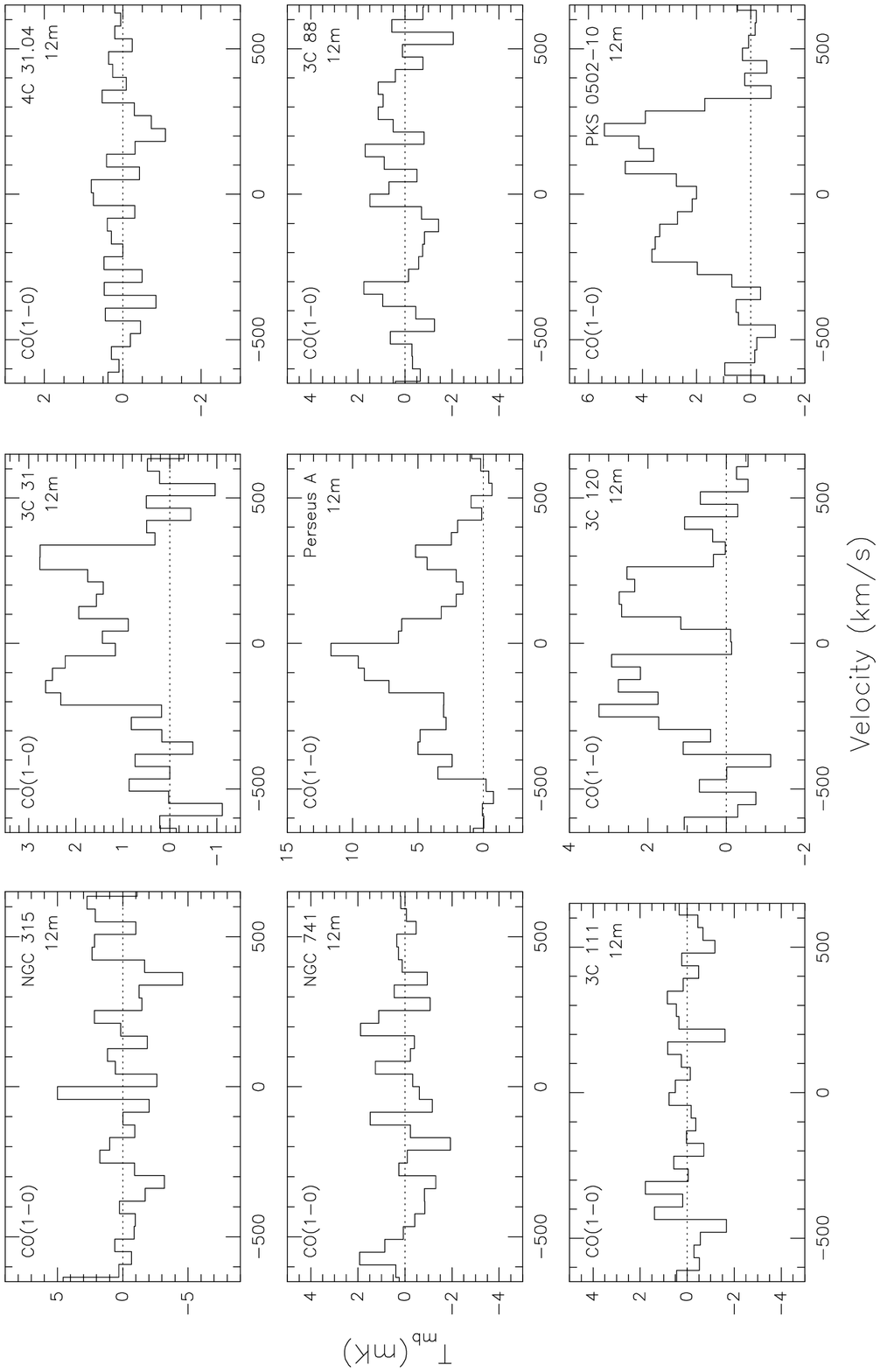}{9.0 in}{0}{90}{90}{-270}{-20}
\figurenum{1a}
\caption{}
\end{figure}

\begin{figure}[h]
\plotfiddle{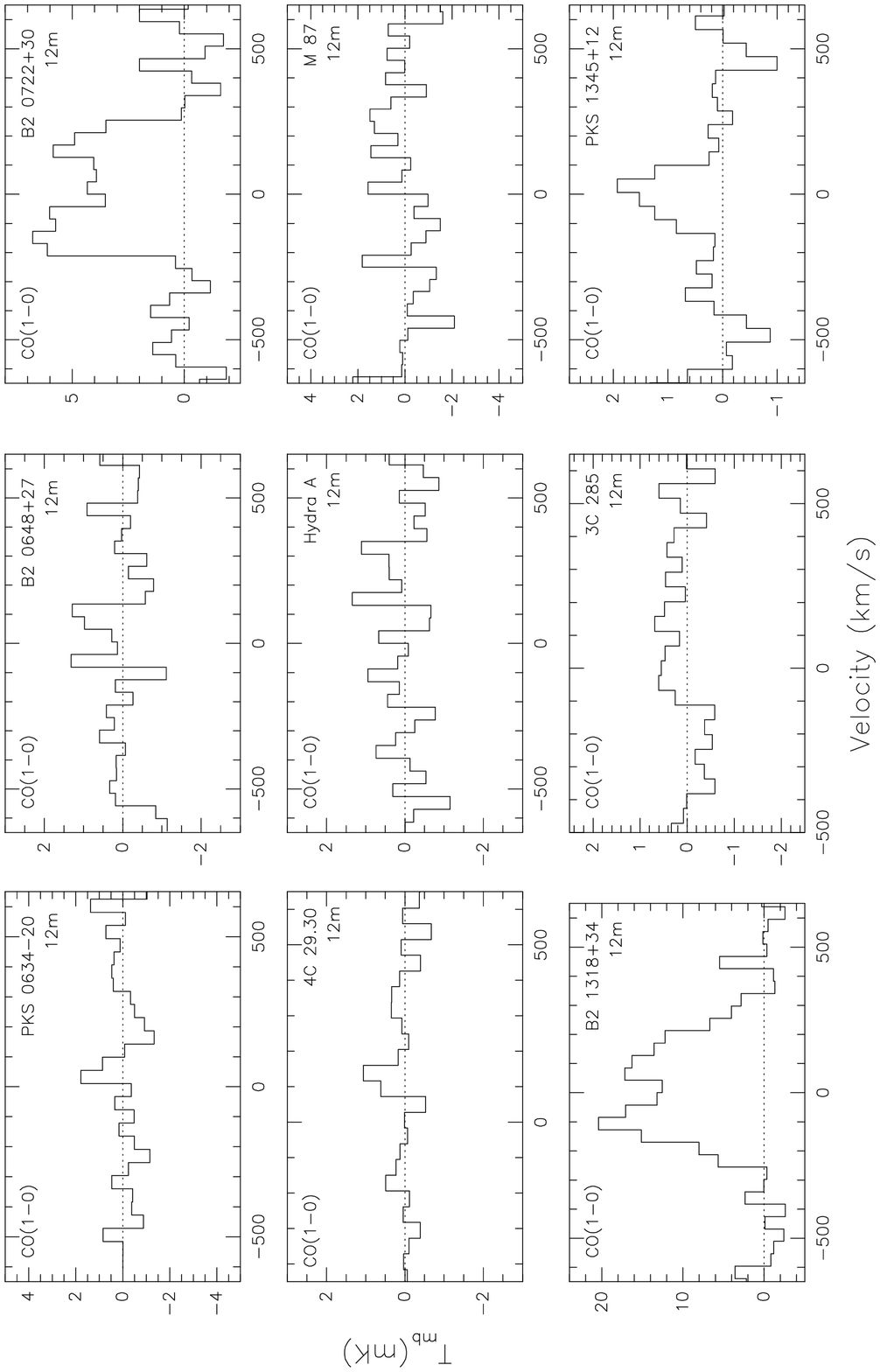}{9.0 in}{0}{90}{90}{-270}{-20}
\figurenum{1a}
\caption{}
\end{figure}

\begin{figure}[h]
\plotfiddle{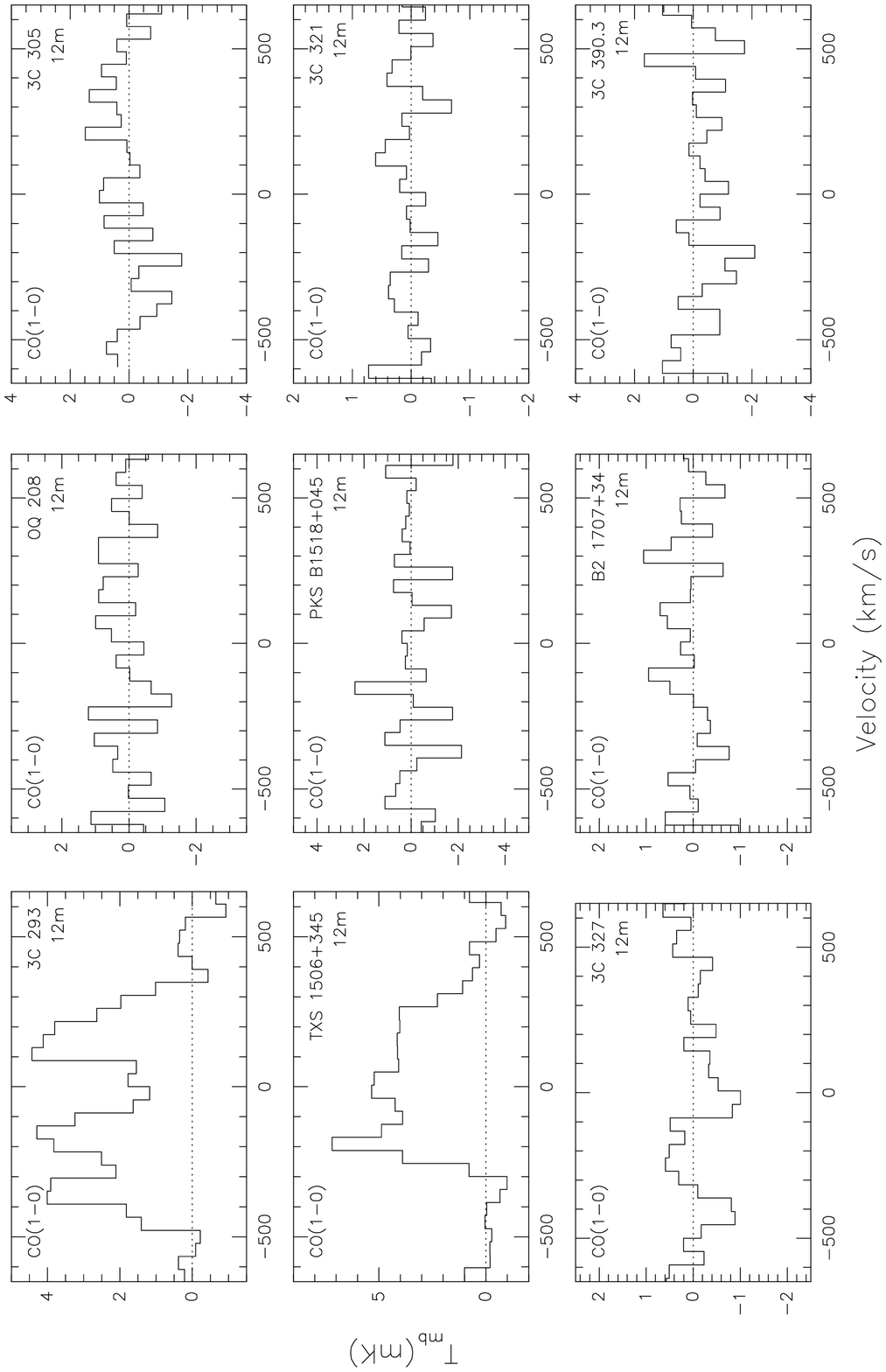}{9.0 in}{0}{90}{90}{-270}{-20}
\figurenum{1a}
\caption{}
\end{figure}

\begin{figure}[h]
\plotfiddle{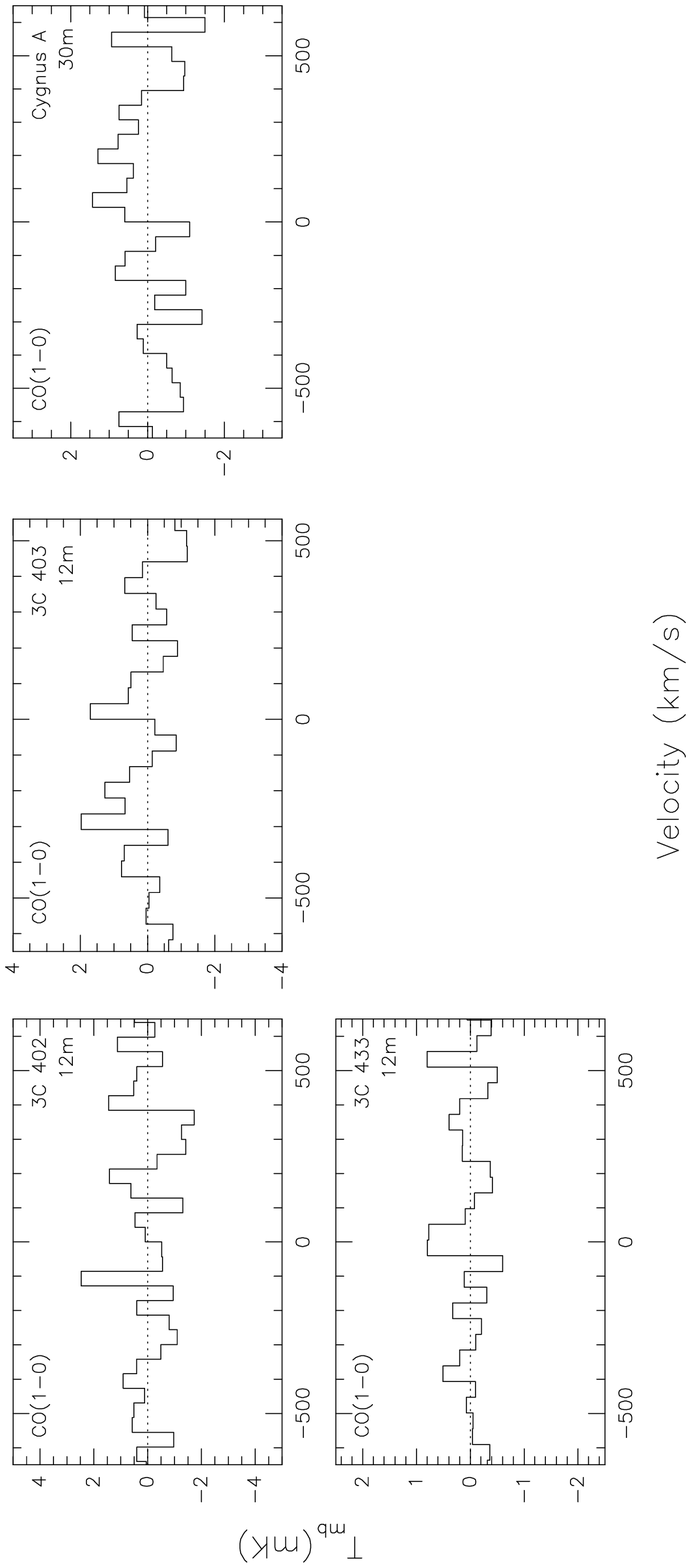}{9.0 in}{0}{90}{90}{-270}{-20}
\figurenum{1a}
\caption{}
\end{figure}

\begin{figure}[h]
\plotfiddle{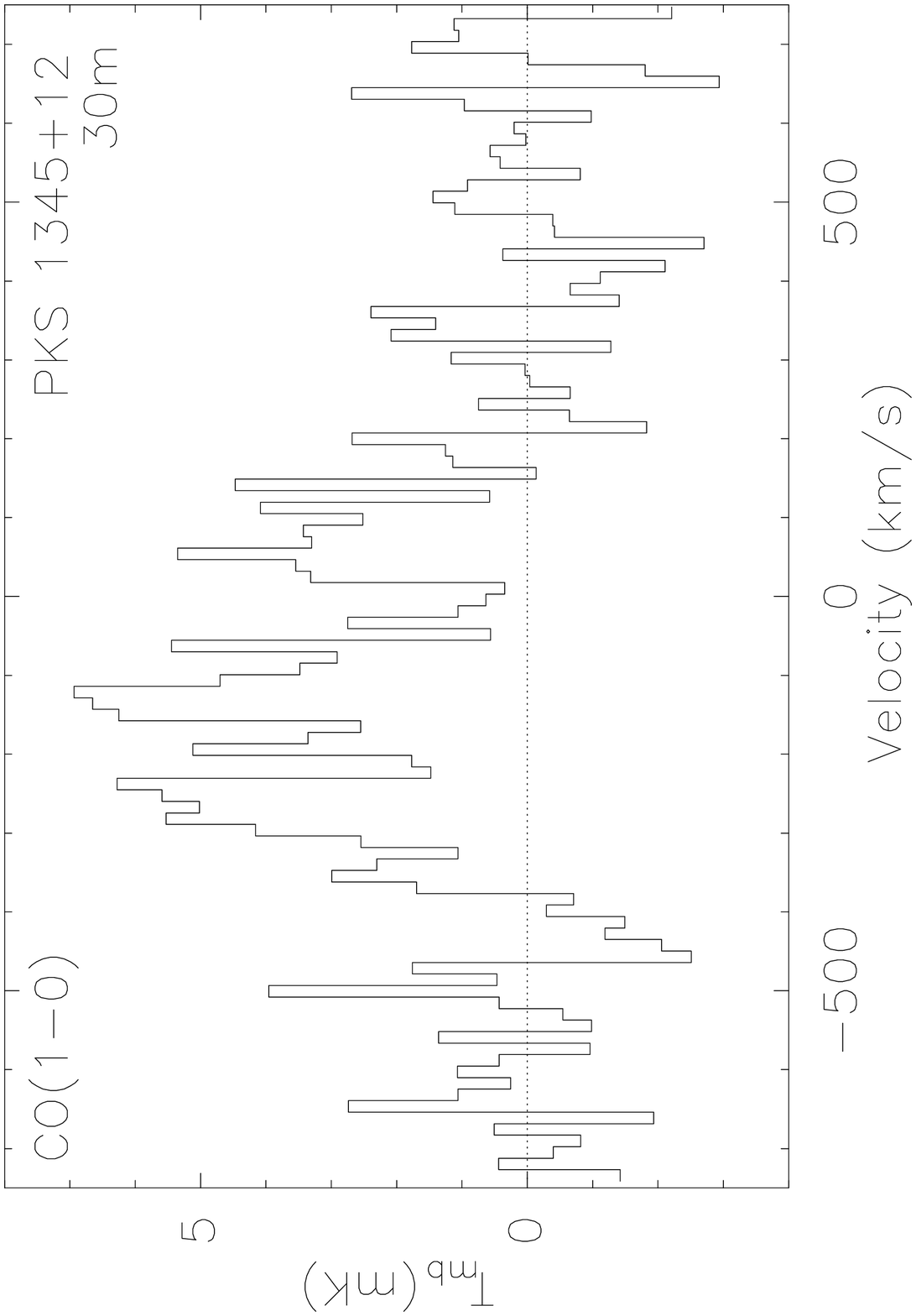}{9.0 in}{0}{90}{90}{-270}{-20}
\figurenum{1b}
\caption{}
\end{figure}

\begin{figure}[h]
\plotfiddle{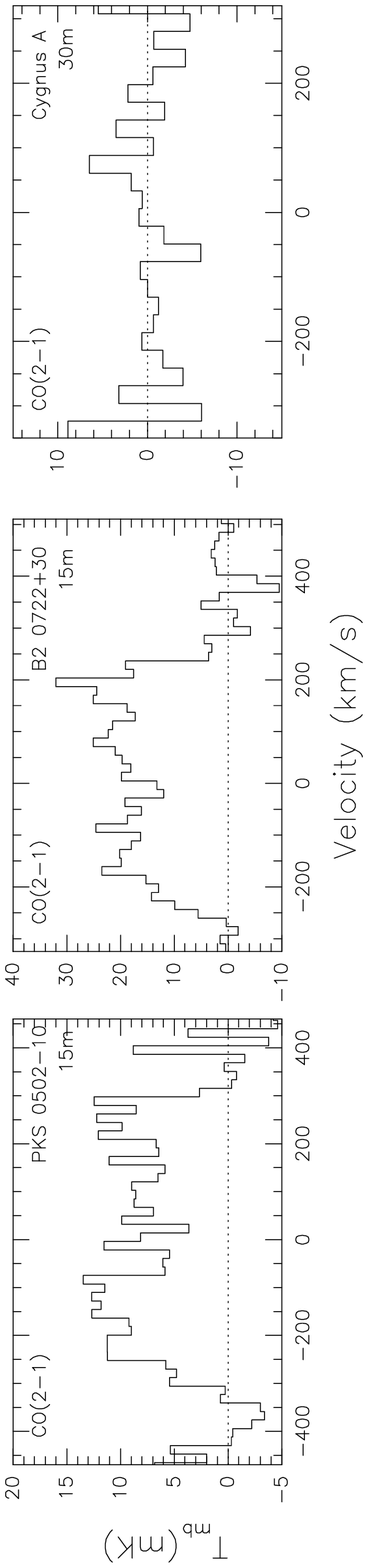}{9.0 in}{0}{90}{90}{-270}{0}
\figurenum{2}
\caption{}
\end{figure}

\begin{figure}[h]
\plotfiddle{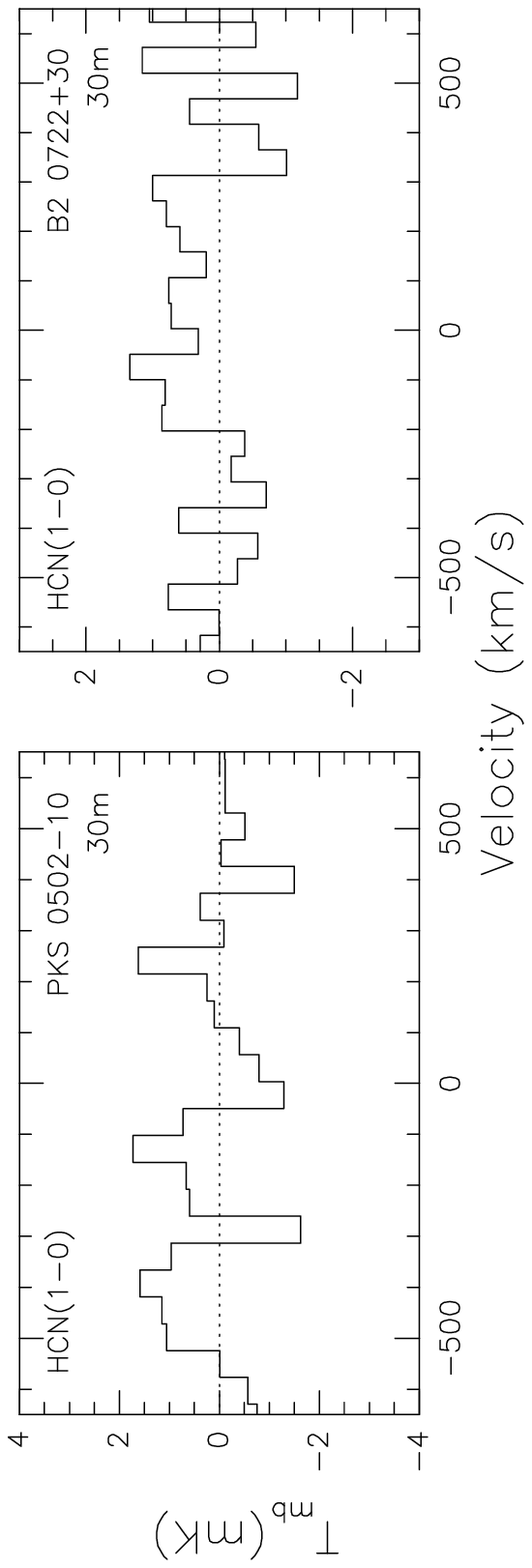}{9.0 in}{0}{90}{90}{-270}{0}
\figurenum{3}
\caption{}
\end{figure}


\begin{figure}[h]
\plotfiddle{Evans.fig5.ps}{9.0 in}{0}{90}{90}{-270}{0}
\figurenum{5}
\caption{}
\end{figure}

\begin{figure}[h]
\plotfiddle{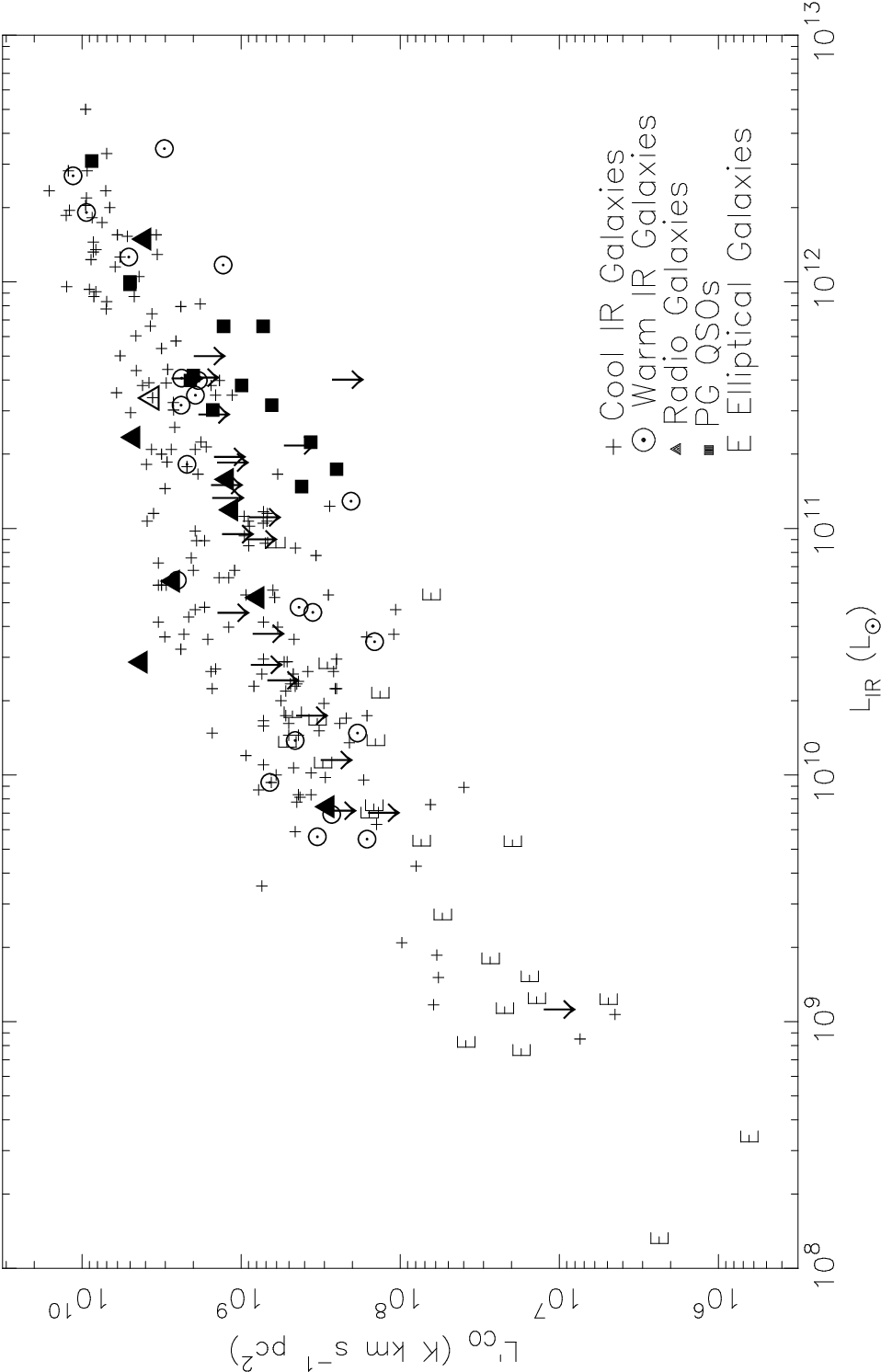}{9.0 in}{0}{90}{90}{-270}{0}
\figurenum{6a}
\caption{}
\end{figure}

\begin{figure}[h]
\plotfiddle{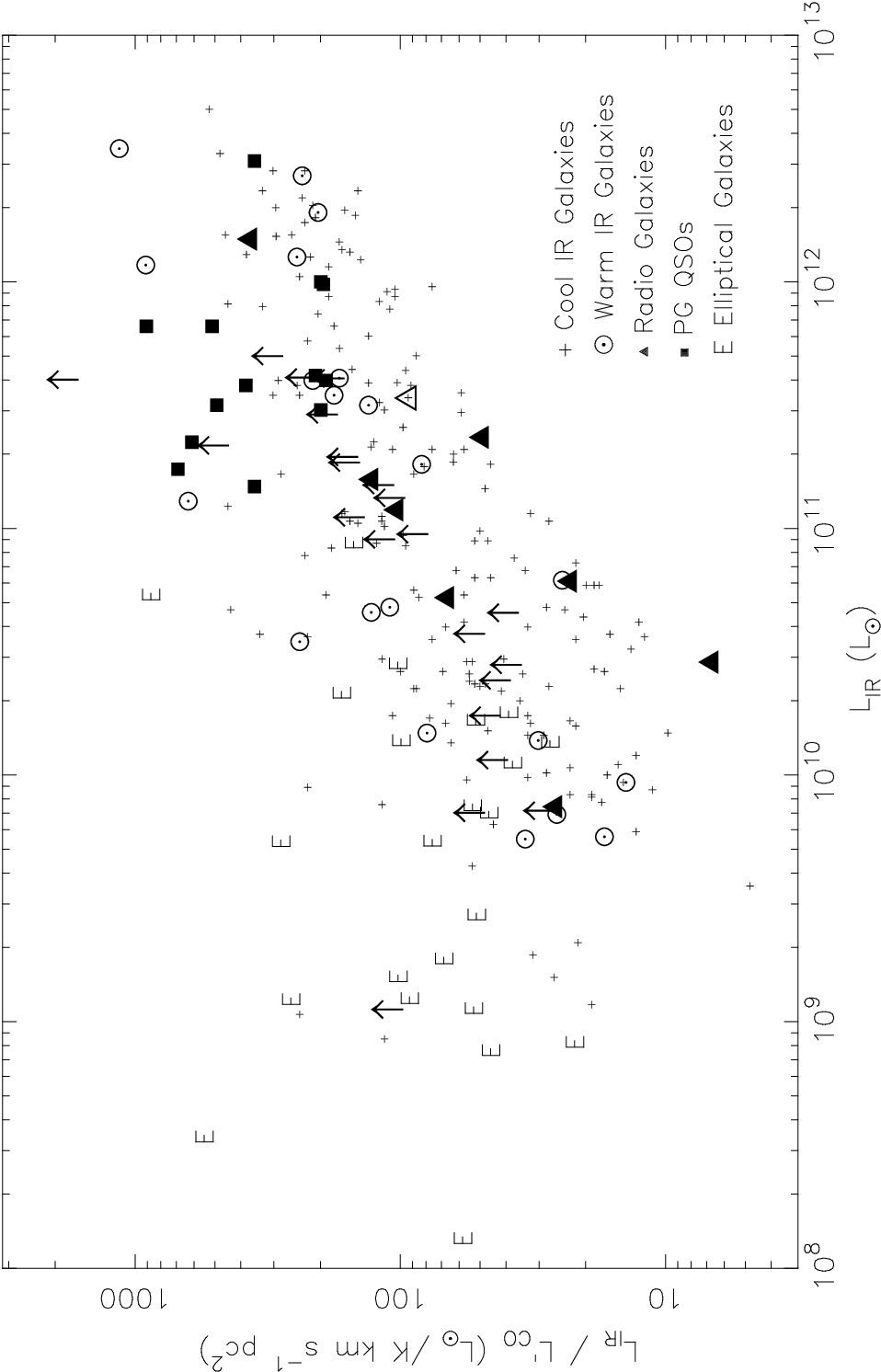}{9.0 in}{0}{90}{90}{-270}{0}
\figurenum{6b}
\caption{}
\end{figure}

\begin{figure}[h]
\plotfiddle{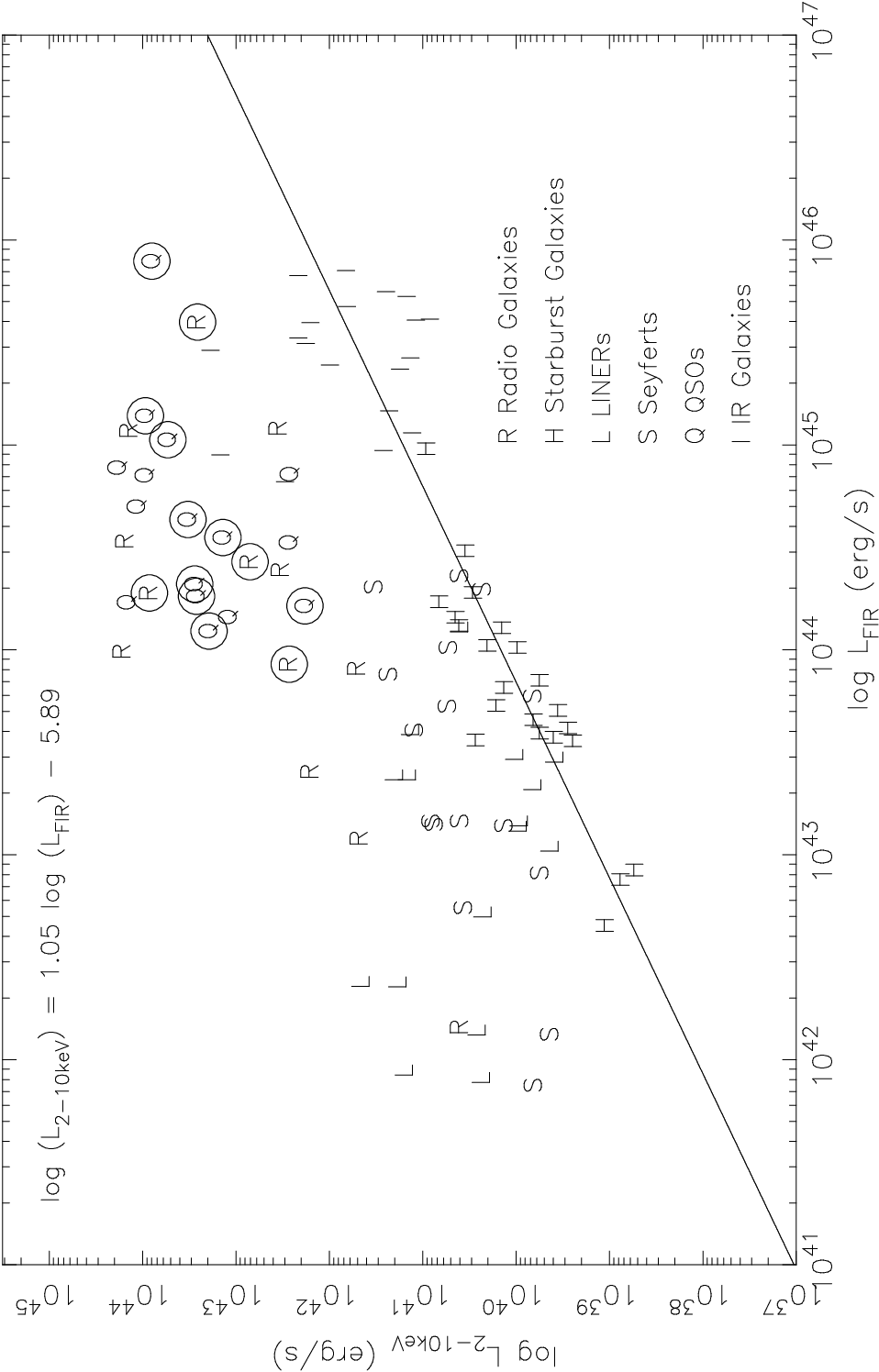}{9.0 in}{0}{90}{90}{-270}{0}
\figurenum{7a}
\caption{}
\end{figure}

\begin{figure}[h]
\plotfiddle{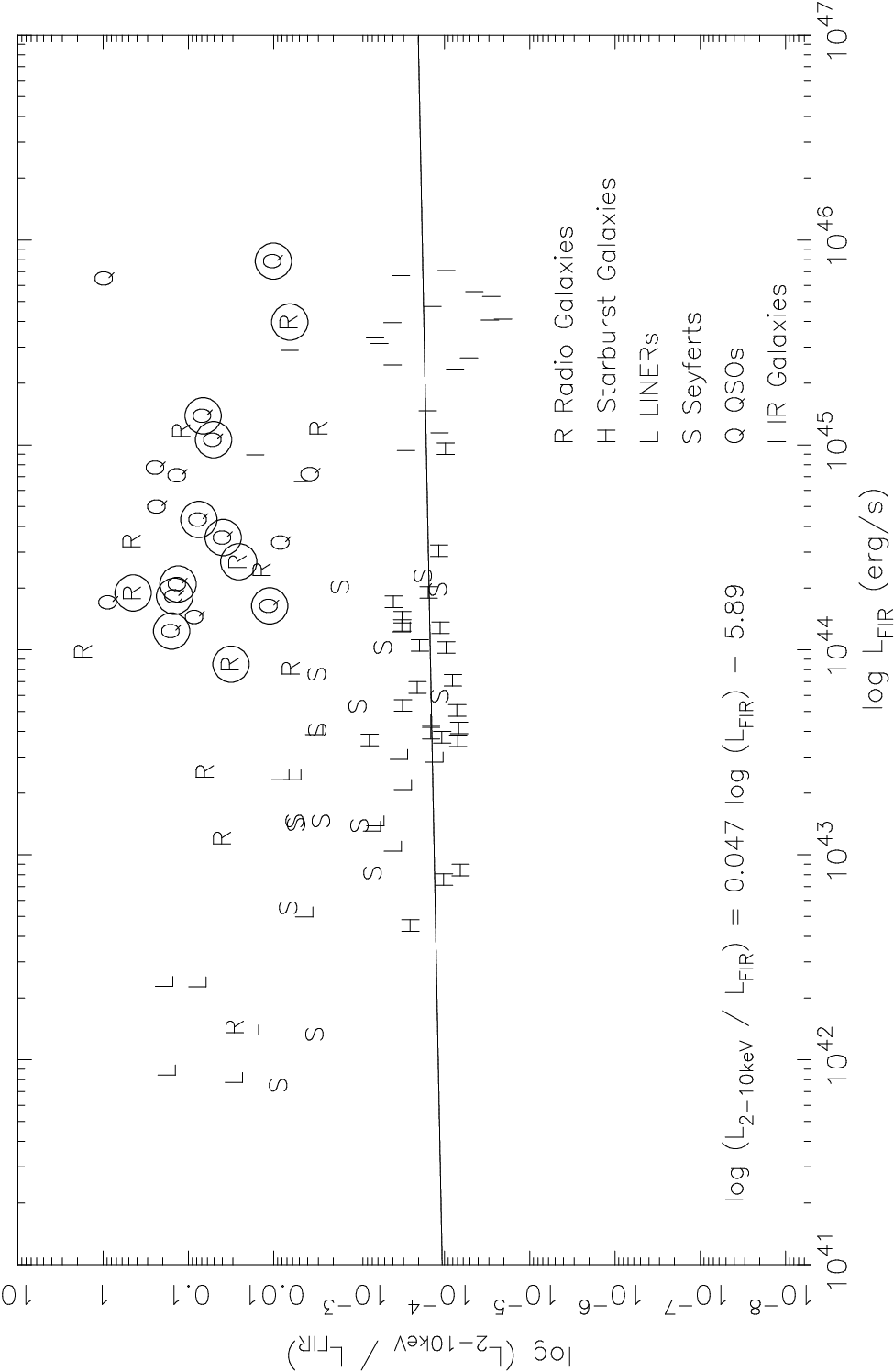}{9.0 in}{0}{90}{90}{-270}{0}
\figurenum{7b}
\caption{}
\end{figure}

\begin{figure}[h]
\plotfiddle{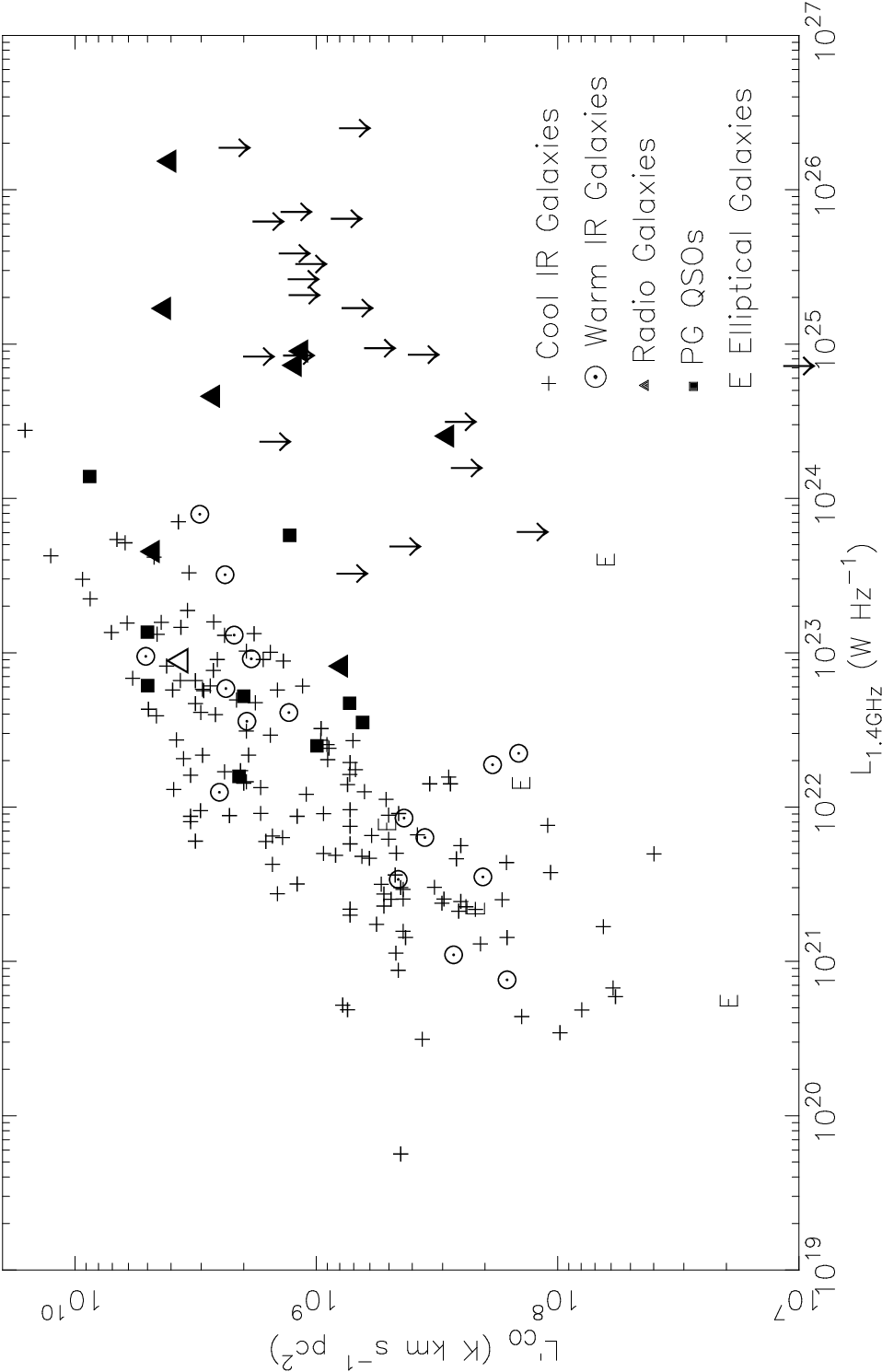}{9.0 in}{0}{90}{90}{-270}{0}
\figurenum{8a}
\caption{}
\end{figure}

\begin{figure}[h]
\plotfiddle{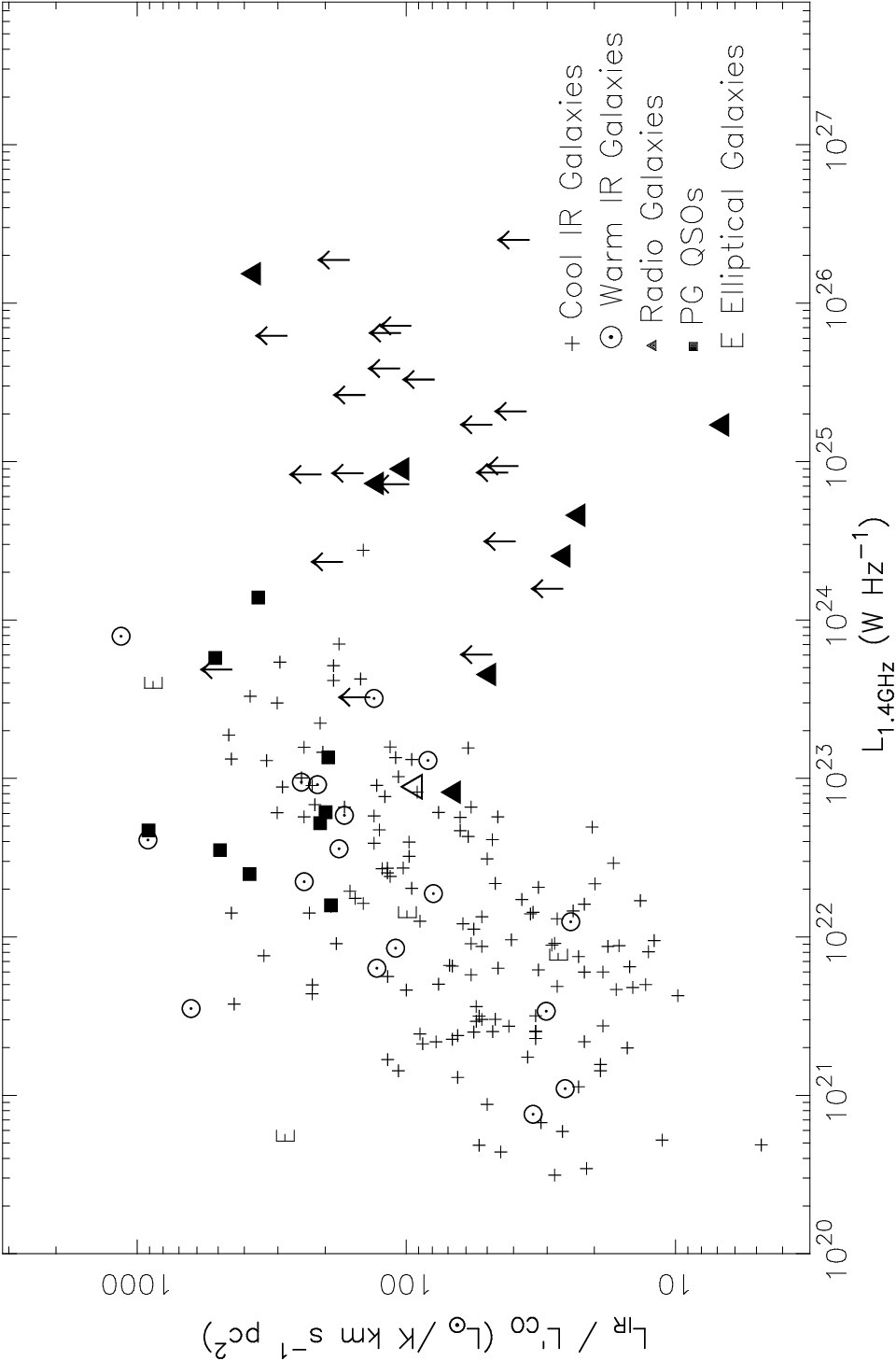}{9.0 in}{0}{90}{90}{-270}{0}
\figurenum{8b}
\caption{}
\end{figure}

\end{document}